\documentclass[11pt,english,a4paper,floatfix]{article}
\usepackage{jcappub}
\usepackage{multirow}
\usepackage[normalem]{ulem}
\pdfoutput=1
\usepackage{bm}
\usepackage{amsfonts}
\usepackage[utf8]{inputenc}
\graphicspath{{Figures/}}

\usepackage[dvipsnames]{xcolor}
\usepackage{graphicx}
\usepackage{amssymb}
\usepackage{amsmath}
\usepackage{times} 
\usepackage{multirow}
\usepackage{url}
\usepackage{cuted}
\usepackage[english]{babel}
\usepackage{enumerate}
\usepackage[normalem]{ulem}
\usepackage{enumitem}
\usepackage{multirow}
\usepackage{float}
\usepackage{booktabs}
\usepackage{makecell}
\usepackage{placeins}
\usepackage{newtxtext,newtxmath}
\usepackage[T1]{fontenc}
\usepackage{amsmath}	
\usepackage{csquotes}
\usepackage[dvipsnames]{xcolor}
\usepackage[dvipsnames]{xcolor}

\newcommand{\Rtwo}{SH0ES'22 }

\begin{document}
	
	
	\title{Effects of a local physics change on the SH0ES determination of $H_0$ }
	
	\author[a, 1]{Ruchika \note{Corresponding author.}}
        \author[b]{Leandros Perivolaropoulos,}
        \author[a]{Alessandro Melchiorri}

	\affiliation[a]{Physics Department and INFN, Universit\`a di Roma ``La Sapienza'', Ple Aldo Moro 2, 00185, Rome, Italy}	
	\affiliation[b]{Department of Physics, University of Ioannina, GR-45110, Ioannina, Greece}

	\emailAdd{ruchika.ruchika@roma1.infn.it}
        \emailAdd{leandros@uoi.gr}
        \emailAdd{alessandro.melchiorri@roma1.infn.it}


\abstract{
The Hubble tension, a significant discrepancy between the Hubble constant ($H_0$) values derived from early-time (Cosmic Microwave Background and Baryon Acoustic Oscillations) and late-time (Cepheid-calibrated Type Ia Supernovae) measurements, remains a major challenge in cosmology. Traditional attempts to resolve this tension have struggled to maintain consistency with dynamical and geometrical probes at redshifts $0.01 < z \lesssim 2.5$. We explore a novel model introducing new degrees of freedom in local physical laws affecting calibrators like Cepheids and Type Ia Supernovae within a distance of $d \lesssim 50$ Mpc ($z \lesssim 0.01$). Specifically, we incorporate a gravitational transition causing a change in the gravitational constant ($G$) at a specific distance, affecting the Cepheid Period-Luminosity Relation (PLR) and the absolute magnitude of SNe Ia. We verify the inverse scaling of SN luminosity $L$ with Chandrasekhar Mass $M_C$ in a changed $G$ scenario as predicted using a semi-analytical model in a recent theoretical study \cite{Wright2018}. Fixing $\Delta G/G \approx 0.04$, our model naturally resolves the Hubble tension, yielding a best-fit $H_0$ value consistent with the Planck measurement, even without using Planck data. This approach suggests a potential resolution to the Hubble tension by aligning $H_0$ with high-redshift CMB measurements. }

\maketitle



\section{Introduction}

The Hubble tension, a discrepancy of around \(5\sigma\) between the Hubble constant (\(H_0\)) values derived from early-universe observations, such as the Cosmic Microwave Background (CMB) and Baryon Acoustic Oscillations (BAO), and late-universe local measurements, using Cepheid-calibrated Type Ia Supernovae (SNe Ia), represents one of the most significant challenges in modern cosmology. The standard cosmological model, \(\Lambda\)CDM, yields an \(H_0\) value of \(67.66 \pm 0.42 \text{ km/s/Mpc}\) from high-redshift CMB measurements \cite{Planck:2018vyg}. In stark contrast, local universe measurements by the SH0ES collaboration report an \(H_0\) value of \(73.04 \pm 1.04 \text{ km/s/Mpc}\) from Cepheid-calibrated SNe Ia \cite{Riess:2021jrx}. Additionally, another local universe measurement using the tip of the red-giant branch (TRGB) calibrated SNe Ia produces an \(H_0\) value of \(69.8 \pm 0.8 (\text{stat}) \pm 1.7 (\text{sys}) \text{ km/s/Mpc}\) \cite{Freedman:2019jwv,Yuan:2019npk}, which is within \(2\sigma\) of both aforementioned measurements. Interestingly, TRGB calibration using GAIA DR3 parallax measurements of \(\omega-\text{Centauri}\) results in an \(H_0\) value of \(72.1 \pm 2.0 \text{ km/s/Mpc}\) \cite{Soltis:2020gpl,Freedman:2020dne}.

Numerous studies have sought to identify unknown systematic effects in local universe measurements that might explain this Hubble discrepancy \cite{Efstathiou:2013via,Cardona:2016ems,Zhang:2017aqn,Feeney:2017sgx,Dhawan:2017ywl,Follin:2017ljs,Riess:2018kzi,Shanks:2018rka,Freedman:2019jwv,Huang:2019yhh,Yuan:2019npk,Soltis:2020gpl,Freedman:2020dne,Riess:2020fzl}. However, the consistency of this discrepancy across a wide array of probes suggests that systematic errors alone are unlikely to account for the Hubble tension.

In response, many proposed solutions have emerged, typically divided into three main categories: early-time models, late-time models (H(z) deformation), and ultralate local physics transitions or systematics.

\textit{Early-time models} introduce modifications to the standard cosmology at or before the recombination epoch. These models aim to decrease the sound horizon scale at recombination through mechanisms such as Early Dark Energy (EDE), modified gravity , or dark radiation \cite{Poulin:2018cxd,Evslin:2017qdn}. By reducing the sound horizon scale, these models can increase the inferred value of \(H_0\) while maintaining consistency with the CMB angular scale measurements. However, despite fine-tuning, these models only reduce the statistical significance of the Hubble tension without fully resolving it. They also tend to favor a higher value of the matter density parameter \(\Omega_m\), exacerbating the \(S_8\) tension \cite{DiValentino:2021izs}.

\textit{Late-time models (H(z) deformation)} assume a deformation of the Hubble expansion history \(H(z)\) at late cosmological times (\(z \lesssim 2\)). These models adjust the expansion rate to reconcile the \(H_0\) values derived from the sound horizon scale with those obtained from the distance ladder method \cite{Bernal:2016gxb,Verde:2019ivm,Knox:2019rjx,Dutta:2018vmq}. However, these deformations are highly constrained by BAO and SNe Ia data at redshifts larger than \(z \approx 0.1\), making it challenging to achieve a consistent fit across all datasets. The \(\Lambda s\)CDM model, which introduces a high-redshift transition in \(H(z)\), shows some promise but still struggles with intermediate/low redshift BAO data \cite{Akarsu:2021fol}.

\textit{Ultralate local physics transitions or systematics} suggest changes in the physical laws or environmental conditions affecting Cepheid-calibrated SNe Ia between the second and third rungs of the distance ladder. This approach posits that \(M_B\) in the Hubble flow rung (\(z \in [0.01, 0.1]\)) is lower than in the calibration rung, leading to a lower inferred \(H_0\). While this model is testable and may explain some observed discrepancies, it suffers from fine-tuning and lacks a clear theoretical motivation for the transition at such low redshifts \cite{Marra:2021fvf}.

Previous studies have investigated local physics transitions as a solution to the Hubble tension. Perivolaropoulos and Skara \cite{Perivolaropoulos:2022khd} reanalyzed the SH0ES data, keeping \(M_W^H\) constant and allowing for different SNe Ia absolute magnitudes (\(M_{B1}\) and \(M_{B2}\)) before and after the transition. They found that this approach could partially alleviate the tension but did not fully resolve it. Wojtak and Hjorth \cite{Wojtak:2022bct} examined the color parameter \(\beta\) of SNe Ia and found significant differences between the calibration and cosmological samples, suggesting that the local measurement of \(H_0\) depends on the choice of SN reference color. By adjusting the reference color, they could reconcile the local \(H_0\) value with the Planck 
measurement. Another interesting study by Wright and Li~\cite{Wright2018} while using a semi-analytic model for SNe light curves found inverse scaling of Chandrasekhar Mass ($M_C$) with Luminosity of Type I-a supernovae which is contradictory to standard $L \propto M_c$ SN relation.

Building on these insights and motivated by \cite{Perivolaropoulos:2022khd}, Ruchika et al. \citep{Ruchika:2023ugh} conducted an analysis adopting the local physics transition approach.  Specifically, they explore a sharp transition in the gravitational constant (\(G\)) in the very late universe and its potential to resolve the Hubble tension by modifying the local distance ladder's physics. Their analysis introduced two novel features compared to previous studies:
1. It incorporated the physics of a gravitational transition, enforcing a physically motivated transition of the Cepheid Period-Luminosity Relation (PLR) intercept at the same distance (time) as the transition of the SNe Ia absolute magnitude.
2. It enforced this transition on \(M_W^H\) rather than simply allowing it, providing a more natural assumption in the context of a gravitational transition. Both \(M_W^H\) and \(M_B\) are expected to change. Future analyses could consider connecting both with \(\Delta G / G\), making \(\Delta G / G\) the only new parameter to fit.
They utilised \Rtwo distances for SN galaxies and did not fit for all the parameters of the distance ladder but only one parameter which is $M_B$ (absolute magnitude of Sne Type Ia).\\
In this paper which is a follow-up paper of \citep{Ruchika:2023ugh}, we fit for period-luminosity relation $M_W^H$ and $b_W$ parameters (slope and intercept parameters) in the anchor box utilising cepheids datasets in the Milky Way, LMC and NGC4258 galaxies. Additionally, we also fit for calibrator galaxies distances  $\mu_i$ using both cepheids and SNe datasets in calibrator galaxies along with the absolute magnitude of SNe ($M_B$ parameter).
 We allow for the SNe Ia standardized peak luminosity to vary with Chandrasekhar mass as $L \propto M_c^n$, where $n$ is the scaling index and is free to take both positive and negative values. We obtained a full posterior distribution and parameter correlation by fitting Cepheids and SNe using the emcee MCMC fitting technique. Using this approach and fixing \(\Delta G / G \simeq 0.04\), which is well consistent with nucleosynthesis constraints, we find that the Hubble tension is naturally and fully resolved. The results obtained lead to a best-fit value of \(H_0\) that is more constrained and significantly more consistent with the Planck value compared to previous studies, even without including a Planck prior or Planck data, as done in the case of Early Dark Energy (EDE) models.

The structure of this paper is as follows: In Section \ref{stdladder}, we describe the method used in our analysis, including the construction of the standard distance ladder and the incorporation of a gravitational transition. Section \ref{sec:data} provides a detailed description of the observational data used, including Cepheids and SNe Ia. In Section \ref{sec:standardfit}, we present the fitting procedure for the distance ladder with a single PLR (no \(G\)-transition) and validate the results. Section \ref{Gtrans-theory} discusses the effects of a \(G\)-transition on the determination of the Hubble constant. In Section \ref{sec:Gtransitionfit}, we fit the distance ladder to a \(G\)-transition and present the results. Section \ref{sec:comparison} compares the models, and finally, Section \ref{sec:discussion} summarizes our findings and discusses future studies.

\section{Hubble Determination by Standard Distance Ladder}\label{stdladder}

In this section, we attempt to determine the value of the Hubble constant (\(H_0\)) by constructing the standard distance ladder, following the methodology employed by the SH0ES team in their analysis. The distance ladder involves three crucial steps to reach cosmological redshifts where the recessional velocities of galaxies are primarily due to cosmic expansion rather than peculiar motion. Such galaxies are referred to as "Hubble flow galaxies."

The first step involves calibrating distances to various local objects to eventually reach cosmological redshifts. For instance, in the Milky Way, distances to Cepheids are measured using parallax methods facilitated by the Hubble Space Telescope (HST), Hipparcos, and Gaia. Once distances and apparent magnitudes are known, the luminosities of these Cepheids can be calculated. By applying the Period-Luminosity Relation (PLR), the luminosities of more distant Cepheids can be inferred. The nearby Cepheids, whose distances can be directly measured, are termed \textit{anchors}, such as those in the Milky Way. Other anchor Cepheids are located in the Large Magellanic Cloud (LMC), Andromeda Galaxy (M31), and NGC4258.

The second step extends the distance ladder to more distant Cepheids. The period and apparent magnitude of these Cepheids can be accurately measured. Using the PLR (assuming it holds true), the luminosities of these Cepheids are estimated. These Cepheids, whose luminosities are determined using the PLR, are termed \textit{calibrators}. The host galaxies of these calibrators are chosen to also host Type Ia Supernovae (SNe Ia), allowing for the calculation of the absolute magnitudes of these supernovae. SNe Ia, well-known as standardizable candles, use this absolute magnitude information to calibrate Hubble flow supernovae, thus extending the distance-redshift relation beyond \(z=1\). This is the third and final step in the distance ladder. The slope of the distance-redshift relation as \(z \rightarrow 0\) provides the value of the Hubble constant \(H_0\).

Here is a qualitative description of the three steps in the distance ladder:
\begin{itemize}
    \item \textbf{Anchor Step:} Calibration of the Cepheid PLR using geometric distances obtained from Water MASERs for the galaxy NGC4258, Detached Eclipsing Binaries (DEBs) for the LMC and M31, and parallaxes for Cepheids in the Milky Way.
    \item \textbf{Calibrator Step:} Using the calibrated Cepheid PLR to infer the distances to SNe Ia host galaxies, which in turn determines the absolute magnitudes of these supernovae.
    \item \textbf{Hubble Flow Supernovae:} Calculating \(H_0\) using the inferred SNe Ia absolute magnitudes and the intercept of the Hubble flow SNe Ia apparent magnitude-redshift relation.
\end{itemize}

The SH0ES distance ladder method involves the following steps:
\begin{enumerate}
    \item \textit{Anchor Step:} The anchor galaxies are located up to approximately 7 Mpc. The calibrated Cepheid PLR is used to infer the distances to SNe Ia hosts with observations of Cepheid variables in each galaxy, which in turn determines the absolute magnitude of these explosions.
    \item \textit{Calibrator Step:} The calibrator galaxies are located between 7 Mpc and 40 Mpc, according to current telescope sensitivities.
    \item \textit{Hubble Flow Step:} \(H_0\) is finally calculated using the inferred SNe Ia absolute magnitude and the intercept of the Hubble flow SNe Ia apparent magnitude-redshift relation. Hubble flow supernovae range from redshifts \(z = 0.023\) to \(z = 0.15\).
\end{enumerate}

Once the standardized magnitude is known, the following relation can be used to infer the value of the Hubble constant \cite{Riess:2016jrr}:
\begin{equation}\label{eq:H0MB}
\log(H_0) = \frac{M_B + 5 \alpha_B + 25}{5}
\end{equation}
where \(\alpha_B\) is the observed intercept of the \(B\)-band apparent magnitude-redshift relation for numerous SNe Ia in the Hubble flow.
\section{Description of observational data used: Cepheids and SNe-Ia}
\label{sec:data}

We will use the same data that \cite{Riess:2021jrx} used in their analysis to fit the distance ladder for both sets of hypotheses. To be explicit, we summarize this dataset below.

For the distances to anchors, the results were taken from the following sources:
\begin{itemize}
    \item For Milky Way Cepheids, we used GAIA EDR3 based parallaxes \cite{GaiaEDR3}. 
    \item For the LMC, we use the distance estimate based on Detached Eclipsing Binaries (DEBs). The physical sizes of the stars in a DEB system can be determined using radial velocity and light curve measurements. These physical sizes can, in turn, be translated into a luminosity distance to the system. Pietrzyński et al. \cite{Pietrzynski:2019jed} found the distance modulus to the LMC to be $\mu = 18.477 \, \pm \, 0.026$. 
    \item For the galaxy NGC4258, we use an improved distance estimate obtained from interferometric observations of water megamasers orbiting around its supermassive black hole \cite{Humphreys:2013eja}. The distance measurement is $D= 7.54\, \pm \, 0.17$ (stat) $\pm$ 0.10 (sys)~Mpc which corresponds to a distance modulus $\mu = 29.387 \, \pm \, 0.057$ \cite{Riess:2016jrr}. 
\end{itemize}

In our MCMC analyses to fit the distance ladder, we will use the "direct" distance measures to the LMC and NGC4258 anchors as Gaussian priors, whereas the distances to MW Cepheids are inferred from the parallaxes. This is explained in further detail in section~\ref{sec:standarddistpriors}.

The observational data for Cepheid periods and apparent magnitudes, as well as SNe Ia apparent magnitudes, were tabulated in several papers by the SH0ES collaboration. We use the following sources, which are also used in the analysis of \cite{Riess:2021jrx}:
\begin{itemize}
    \item Hubble Space Telescope (HST) WFC3-IR photometry and pulsation period of Cepheids in NGC4258 and in the 19 calibrator galaxies were taken from Table 4 of \cite{Riess:2016jrr}. Outlier rejection via a global $2.7\sigma$ clipping of the PLR has already been applied to these Cepheids.
    \item WFC3-IR photometry and pulsation period for the LMC Cepheids was taken from Table 2 of \cite{Riess:2019cxk}.
    \item WFC3-IR photometry, pulsation period, and Gaia EDR3 parallaxes for MW Cepheids were taken from Table 1 of \cite{Riess:2021jrx}\footnote{Our analysis uses the parallaxes given in the $\pi_{EDR3}$ column of Table 1 of SHOES'21 \cite{Riess:2021jrx}, which does not include the residual parallax offset of -14 $\mu$as as identified in \cite{Riess:2021jrx}.}. 64 MW Cepheids out of the given 75 are used. We removed MW Cepheids with unreliable Gaia EDR3 parallaxes and those which are indicated as possible outliers in \cite{Riess:2021jrx}. We also remove the Cepheid S Cru, since its photometry has been transformed from ground-based systems, due to failed HST acquisition. We convert these parallaxes ($\pi$) to distance moduli ($\mu$) using the standard relation $\pi = 10^{-0.2 (\mu - 10)}$. 
    \item Standardized $B$-band photometry for 19 SNe Ia in the 19 calibrator galaxies was taken from Table 5 of \cite{Riess:2016jrr}.
\end{itemize}

We follow \cite{Riess:2016jrr} for converting the HST WFC3-IR photometry to the NIR Wesenheit system. We assume a reddening law with $R_V = 3.3$, which yields $R = 0.386$ \cite{Fitzpatrick:1999by}. The apparent magnitudes corrected for reddening are thus of the form,
\begin{equation}
    m_H^W = m_{F160W} - 0.386 \: (m_{F555W} - m_{F814W}),
\end{equation}
where the subscript $H$ on the left denotes that we are using the H-band HST filter 160W, which is corrected for reddening using the F555W (V) and F814W (I) bands, where the $W$ super/subscripts denote NIR Wesenheit magnitudes. This reddening law is consistent with the procedure used in \cite{Riess:2021jrx}.

For Hubble flow SNe, we are using the value of $a_B$= 0.71273 $\pm$ 0.00176 as determined in \cite{Riess:2016jrr}. This value was determined by using only Hubble flow SNe Ia at redshifts $0.023<z<0.15$. This range lies well beyond the distances to calibrators and thus this value can be used no matter which hypothesis (with/without) $G$-transition we are assuming.


\section{Fitting the distance ladder with single PLR (no $G$-transition)}
\label{sec:standardfit}

We present here the methodology used to fit the standard distance ladder with a single PLR (without assuming a $G$-transition), i.e., the standard scenario. Our procedure is analogous to what is followed by \cite{Riess:2021jrx}, with a few simplifying assumptions.

As done in \cite{Riess:2021jrx}, we also fit the PLR of anchor Cepheids, calibrator Cepheids, and apparent magnitudes of SNe Ia in calibrator galaxies simultaneously. As mentioned in Section \ref{sec:data}, while we use the observed values of the distances to Cepheids in the MW from Gaia, we use priors on the distances to the LMC and NGC4258 from other data sets. The fit yields the distance modulus to the calibrator galaxies as well as the LMC and NGC4258. In addition to these distances, we are also fitting for the Cepheid PLR parameters and the standardized absolute magnitude of SNe Ia (denoted as $M_B$).

This value of $M_B$ can then be used in conjunction with observed apparent magnitudes of Hubble flow SNe to obtain the distances to their host galaxies, and this can be further used to infer the value of the Hubble constant.

For the Cepheids in the anchor galaxies (NGC4258, LMC, and Milky Way), we have the respective expressions for the PLRs,
\begin{equation} \label{eq: N4258_PLR_model}
    m_{H,NGC4258,j}^W = \mu_{NGC4258} + M^W_{H,1} + b_W \left(\log{(P_{N4258,j})} - 1\right),
\end{equation}
\begin{equation} \label{eq: LMC_PLR_model}
    m_{H,LMC,j}^W = \mu_{LMC} + M^W_{H,1} + b_W \left(\log{(P_{LMC,j})} - 1\right),
\end{equation}
\begin{equation} \label{eq: MW_PLR_model}
    m_{H,MW,j}^W = \mu_{MW,j} + M^W_{H,1} + b_W \left(\log{(P_{MW,j})} - 1\right),
\end{equation}
where the observed NIR Wessenheit magnitudes $m^W_H$ on the left are for the $j^{\textrm{th}}$ Cepheid in each calibrator, and the observed periods $P$ of these Cepheids on the right are measured in days. $\mu_{NGC4258}$ and $\mu_{LMC}$ are the distance moduli to NGC4258 and the LMC, respectively, while $\mu_{MW,j}$ denotes the distance modulus to the $j^{\textrm{th}}$ MW Cepheid. For the LMC and NGC4258, the distance moduli represent the average distance to Cepheids in these galaxies. The slope $b_W$ and the intercept $M^{W}_{H,1}$ of the Cepheid PLR are parameters to be determined by fitting the observational data (in this convention, the PLR intercept is defined as the absolute magnitude of a Cepheid with a period $P = 10$ days).

Similarly, for the Cepheids in the calibrator galaxies, we have
\begin{equation} \label{eq: plr_model}
    m_{H,ij}^W = \mu_i + M^W_{H,1} + b_W \left(\log{(P_{ij})} - 1\right),
\end{equation}
where index $j$ labels the Cepheid and index $i$ labels the galaxy. The distance modulus $\mu_i$ to the $i^{\textrm{th}}$ calibrator galaxy is to be determined from our fit.

For calibrator galaxies with the SNe Ia host in them, we have,
\begin{equation} 
\label{eq:SNeIaMB}
    m_{B,i} = \mu_{i} + M_B, 
\end{equation}
where $m_{B,i}$ is the observed $B$-band peak apparent magnitude (after the light curve shape fitting correction type for SN Type I-a) and $\mu_i$ is the distance to the $i^{th}$ calibrator galaxy (already fitted by Cepheids in $i^{th}$ host galaxy). Here $M_B$ is the free parameter denoting standardized $B$-band absolute magnitude of Type Ia SN and it is extracted from a fit to the data.

In what follows, we describe the procedure to simultaneously fit the equations for PLR for the anchor Cepheids, along with the SNe Ia standardized peak absolute magnitude $M_B$ in the calibrator galaxies. The value of $M_B$ is assumed to be the same for all SNe Ia. Thus, after fitting the data to obtain constraints on the free parameter $M_B$, we can use this result in eq.~\ref{eq:H0MB} to infer the constraints on the Hubble constant.

\subsection{Parameter Priors}
\label{sec:standarddistpriors}

For the distance modulus parameters to the LMC and NGC4258, we use Gaussian priors based on the values and 1-$\sigma$ errors quoted in the references discussed in Section~\ref{sec:data}. For the distance modulus parameters for MW Cepheids, we use the mean value calculated from the GAIA EDR3 parallaxes~$\pi_{\textrm{EDR3}}$ corresponding to each Cepheid. The error on the parallax distance measurement is taken into account when we define our likelihoods.

We use uniform priors over a wide range on the unknown parameters of the distance ladder—the slope and intercept of the PLR ($b_W$, $M^W_{H,1}$), the distance moduli to the 19 calibrator galaxies ($\mu_i$), and the standardized SNe Ia peak absolute magnitude ($M_B$). The set of all parameters that we are fitting for, along with the type and range for their priors, are given in Table~\ref{table:basic_prior}.

\begin{table}
\centering
\begin{tabular}{|p{2.1cm} |p{3.1cm}|p{7.0cm}|}
\Xhline{\arrayrulewidth}
\textbf{Parameter}  & \textbf{Prior } & \textbf{Description}\\ 
\Xhline{\arrayrulewidth}
\vspace{1.5cm}
  & \textbf{Uniform } & \\ 
\Xhline{\arrayrulewidth}
$b_W$ &  [-40, 10] & PLR slope\\
$M^W_{H,1}$ &  [-40, 10] & PLR intercept\\
$\mu_i$ &  [1, 50] & distance moduli to the 19 calibrator galaxies\\
$M_B$ & [-21,-17] & SNe Ia absolute magnitude\\
\Xhline{\arrayrulewidth} 
\vspace{1.5cm}
  & \textbf{Gaussian ($\mathcal{N}(\mu,\,\sigma^{2})$) } & \\ 
  \Xhline{\arrayrulewidth} 
$\mu_{\rm LMC}$ & $\mathcal{N}(18.477,0.026^{2} )$ & LMC distance modulus\\
$\mu_{\rm NGC4258}$ & $\mathcal{N}(29.387,0.057^{2} )$ & N4258 distance modulus\\
\Xhline{\arrayrulewidth}  
\end{tabular} 
\caption{This table includes a list of parameters used in the fitting of the standard distance ladder and their corresponding priors. Uniform priors are chosen for the Cepheid PLR slope and intercept, the distance moduli to the calibrator galaxies, and the standardized SNe~Ia absolute magnitude. Gaussian priors are used for distances to the anchor galaxies LMC and NGC4258.}
\label{table:basic_prior}
\end{table}

\subsection{Fit using $\chi^2$ minimisation}
\label{sec:chi2_standard}

We simultaneously fit for the Cepheid period-luminosity empirical relation for anchor (eqs.~\ref{eq: N4258_PLR_model}, \ref{eq: LMC_PLR_model}, and \ref{eq: MW_PLR_model}) and calibrator galaxies (eq.~\ref{eq: plr_model}), and for SNe Ia standardized peak absolute magnitude in calibrator galaxies (eq.~\ref{eq:SNeIaMB}).

The parameters that we are fitting for are the Cepheid PLR intercept and slope parameters ($M^W_{H,1}$ and $b_W$), the distance moduli $\mu_i$ to 19 calibrator galaxies, the distance moduli $\mu_{LMC}$ and $\mu_{NGC4258}$, and the SNe Ia absolute magnitude ($M_B$). Thus, in total there are 24 parameters that we are fitting for.

We first define a $\chi^{2}$ or equivalently a log-likelihood (where $\chi^{2}= -2 \log\mathcal{L}$) and perform a minimization over all the parameters using the publicly available code emcee hammer \cite{ForemanMackey:2012ig} to generate MCMC (Markov Chain Monte Carlo) chains to obtain the best-fit parameter values.

Our total log-likelihood is defined using two contributions, one from Cepheid data and the other from SNe data. For the Cepheids in the anchor and calibrator galaxies, we define the likelihood function as follows:
\begin{equation}
\label{eq:Lcep}
   \log \mathcal{L}_{\textrm{cep}} = -\frac{1}{2} \sum_{ij}  \frac{\left[(m_{H,ij}^W)^{\rm{obs}} - (m_{H,ij}^W)^{\rm{model}} \right]^2}{\sigma_{ij}^2},
\end{equation}
where the index $j$ labels the Cepheids and index $i$ labels the galaxies. The superscript \enquote{obs} corresponds to the observed value of $m_H^W$ and the superscript \enquote{model} corresponds to the value of $m_H^W$ calculated from the theoretical model of the standard distance ladder. The errors $\sigma_{ij}^2$ are taken to be $\sigma^2_{m_{H,ij}^W}$, i.e., the 1-$\sigma$ standard deviation error on the observed~$m_{H,ij}^W$. However, for MW Cepheids, the observable should be thought of as $\left (m_{H,MW,j}^W - \mu_{MW,j} \right)$, thus for these Cepheids we take the error to be the 1-$\sigma$ standard deviation errors on the observed~$m_{H,MW,j}^W$ and the 1-$sigma$ error on $\mu_{MW,j}$ from Gaia parallax measurements, added in quadrature.

For SNe Ia present in the calibrator step, we define the likelihood function as follows:
\begin{equation}
\label{eq:LSN}
   \log \mathcal{L}_{\textrm{SN Ia}} = -\frac{1}{2} \sum_{i} \left[ \frac{(m_{B,i})^{\rm{obs}} - (m_{B,i})^{\rm{model}}}{\sigma_{m_{B,i}}}  \right]^2,
\end{equation}
where $m_{B,i}$ is the corrected $B$-band peak apparent magnitude of the SNe Ia in the $i^{th}$ calibrator galaxy. Again, the superscript \enquote{obs} corresponds to the observed value of $m_{B,i}$ and the superscript \enquote{model} corresponds to the value of $m_{B,i}$ calculated from the theoretical model of the standard distance ladder (equation \ref{eq:SNeIaMB}). $\sigma_{m_{B,i}}$ is the 1-$\sigma$ standard deviation error on the observed $(m_{B,i})^{\rm{obs}}$.

The total likelihood that we use when fitting the Cepheids in the anchor and calibrator galaxies and SNe Ia in only the calibrator galaxies, is the sum of $\mathcal{L}_{\textrm{cep}}$ and $\mathcal{L}_{\textrm{SN Ia}}$,
\begin{equation}
\label{eq:Ltotal}
    \mathcal{L} = \mathcal{L}_{\textrm{cep}} + \mathcal{L}_{\textrm{SN Ia}}.
\end{equation}

Extremizing the final likelihood function $\mathcal{L}$, we obtain the best fit values and 1-$\sigma$ confidence intervals on all the model parameters.

\subsection{Results and validation for fit to the distance ladder}
\label{sec:resultsnoG}

After performing the fit described above, we obtained a minimum $\chi^2$ of 1963.19 for 24 parameters. With 1267 data points, this results in a $\chi^2$ per degree of freedom ($\chi^2_{\textrm{dof}}$) of $1.579$.

Our best-fit value for the standardized $B$-band absolute magnitude of SNe Ia is $M_B = -19.23 \pm 0.03$. Substituting this value of $M_B$ into the $H_0-M_B$ relation (eq.~\ref{eq:H0MB}), we infer a value of the Hubble constant $H_0 = 73.73 \pm 1.12$~km/s/Mpc.

In our reanalysis of the standard distance ladder, we made several simplifications compared to the more detailed analysis of \cite{Riess:2021jrx}. Key differences are summarized below:

\begin{itemize}
    \item We have ignored the dependence of the Cepheid PLR on the Cepheid metallicity. The possibility of PLR breaks based on the pulsation period of Cepheids (P > 10 days or P < 10 days) is also ignored, unlike in the main analysis of \cite{Riess:2021jrx}.
    
    \item We do not include the residual parallax offset of -14 $\mu$as for Gaia EDR3 Cepheids, which \cite{Riess:2021jrx} includes. However, \cite{Riess:2021jrx} also provided an analysis where the residual correction was ignored, and the Hubble constant remained consistent. Hence, ignoring this residual correction is not expected to significantly impact our analysis.

    \item \cite{Riess:2016jrr} considered various combinations of anchors, including Cepheids in M31, to check for systematic effects associated with specific anchors. We use their recommended combination for reporting the final best-fit value of the Hubble constant.

    \item We do not consider the background covariance matrix for Cepheids and SNe Ia. Off-diagonal covariances were also neglected in \cite{Riess:2021jrx}, although these covariances were considered in the later analysis by \cite{Riess:2022mme}.

    \item We use observational datasets from \cite{Riess:2021jrx}, which include 19 SN-Ia host galaxies. We do not incorporate results from \cite{Riess:2022mme}, which include 42 SNe Ia in 37 host galaxies. Although using the SH0ES'22 datasets could enhance our analysis precision, this is not the primary objective of our current research. Therefore, we adhere to a conservative approach and use the \cite{Riess:2021jrx} datasets while conducting the full emcee MCMC analysis to minimize computational cost. 
\end{itemize}

The value of the Hubble constant we infer is consistent with that of \cite{Riess:2021jrx}. We also check our fit parameters for consistency with the results from \cite{Riess:2021jrx}. This comparison is shown for the Cepheid PLR parameters and the standardized SN peak absolute magnitude in Table \ref{table:result_basic_std}. All parameters are in good agreement with the results of SH0ES'21 \cite{Riess:2021jrx}.

\begin{table}
    \centering
\scalebox{0.85}{
\renewcommand{\arraystretch}{2.1}
\begin{tabular}{|c|c|c|c|}
\hline
\hline
Parameter & Our fit & SH0ES'21 & SH0ES'22 \\ 
\hline
\hline
$M^W_H$ & $-5.89 \pm 0.01$ & $-5.92 \pm 0.030$ & $-5.894 \pm 0.018$ \\ \hline
$b_W$  & $-3.28 \pm 0.01$ & $-3.28 \pm 0.06$ & $-3.298 \pm 0.015$ \\ \hline
$M_{B}$ & $-19.23 \pm 0.03$ &  $-19.24 \pm 0.04$ & $-19.253 \pm 0.029$ \\ \hline
$H_0$ (km/s/Mpc) & $73.73 \pm 1.12$ & $73.0 \pm 1.4$ & $73.04 \pm 1.04$ \\ \hline
\hline
\end{tabular}}
\caption{Comparison between the best-fit value and 1-$\sigma$ error on various parameters of the standard distance ladder obtained from our fit and the \cite{Riess:2021jrx} and \cite{Riess:2022mme} fits. Our analysis has tried to emulate that of \cite{Riess:2021jrx} and we find good consistency with both SH0ES fits despite making various simplifications in the formulation of the standard distance ladder. These simplifications lead to minor differences between our parameter estimates and those of the SH0ES team.}
\label{table:result_basic_std}
\end{table}

The other fit parameters are not explicitly mentioned in \cite{Riess:2021jrx}. However, we can compare our fit parameters to the distances to the 19 calibrator galaxies obtained in the earlier reported analysis of \cite{Riess:2016jrr}, which did not include GAIA parallaxes to MW Cepheids. This comparison is tabulated in Table \ref{table:result_basic_dis} and shown in Fig.~\ref{fig:mu_compare_both}, and once again we find good agreement with this study.

\begin{table}
    \centering
\scalebox{0.85}{
\renewcommand{\arraystretch}{1.5}
\begin{tabular}{|c|c|c|c|c|c|c|c|c|}
\hline
\hline
S. No. & Galaxy & Our fit & SH0ES'16 & S. No. & Galaxy & Our fit & SH0ES'16 \\ 
\hline
\hline
1 & M101 & $29.05 \pm 0.02$ & $29.135 \pm 0.045$ & 11 & N3982 & $31.67 \pm 0.05$ & $31.737 \pm 0.069$ \\ 
2 & N4424 & $30.80 \pm 0.10$ & $31.080 \pm 0.292$ & 12 & N5584 & $31.76 \pm 0.03$ & $31.786 \pm 0.046$ \\ 
3 & N4536 & $30.88 \pm 0.04$ & $30.906 \pm 0.053$ & 13 & N3447 & $31.90 \pm 0.03$ & $31.908 \pm 0.043$ \\ 
4 & N1365 & $31.27 \pm 0.05$ & $31.307 \pm 1.057$ & 14 & N3370 & $32.07 \pm 0.04$ & $32.072 \pm 0.049$ \\ 
5 & N1448 & $31.30 \pm 0.03$ & $31.311 \pm 0.045$ & 15 & N5917 & $32.26 \pm 0.08$ & $32.263 \pm 0.102$ \\ 
6 & N4038 & $31.35 \pm 0.08$ & $31.290 \pm 0.112$ & 16 & N3021 & $32.37 \pm 0.06$ & $32.498 \pm 0.090$ \\ 
7 & N2442 & $31.47 \pm 0.04$ & $31.511 \pm 0.053$ & 17 & N1309 & $32.49 \pm 0.04$ & $32.523 \pm 0.055$ \\ 
8 & N7250 & $31.51 \pm 0.06$ & $31.499 \pm 0.078$ & 18 & N1015 & $32.54 \pm 0.06$ & $32.497 \pm 0.081$ \\ 
9 & N4639 & $31.53 \pm 0.05$ & $31.532 \pm 0.071$ & 19 & U9391 & $32.86 \pm 0.05$ & $32.919 \pm 0.063$ \\ 
10 & N3972 & $31.57 \pm 0.06$ & $31.587 \pm 0.070$ & & & & \\ 
\hline
\end{tabular}}
\caption{Comparison between the best-fit value and 1-$\sigma$ error on distances to host galaxies used in the standard distance ladder obtained from our fit and SH0ES'16. Our analysis has tried to emulate that of \cite{Riess:2021jrx} and we find good consistency with SH0ES'16 fits despite making various simplifications in the formulation of the standard distance ladder. These simplifications lead to minor differences between our parameter estimates and those of the SH0ES team.}
\label{table:result_basic_dis}
\end{table}

The most recent analysis of the distance ladder and the Hubble constant determination by the SH0ES team was performed in \cite{Riess:2022mme}. Some of the major improvements in this study were:
\begin{enumerate}
    \item Inclusion of a larger number of calibrators with 42 SNe in 37 host galaxies.
    \item Consideration of correlations between apparent magnitudes of Cepheids in a particular host galaxy.
    \item Improved treatment of the reddening law for Cepheids and its associated uncertainty.
\end{enumerate}

For comparison, the best-fit parameters obtained in \cite{Riess:2022mme} are also shown in Table \ref{table:result_basic_std}. Our fit parameters are consistent with these newer results within $1$-$\sigma$. Thus, our simplified formulation of the standard distance ladder analysis is consistent with the \cite{Riess:2021jrx} analysis in its most important features, and our inferred parameters are also consistent with the more recent results of \cite{Riess:2022mme}.

The neglect of the metallicity corrections in our fit possibly leads us to underestimate the errors, resulting in a large $\chi^2_{\textrm{dof}} \simeq 1.58$, compared to the typical $\chi^2_{\textrm{dof}} \simeq 1.0 - 1.1$ obtained in \cite{Riess:2016jrr}. This difference can be accounted for by a roughly 20-30\% underestimate of the errors due to Cepheid metallicity corrections. In the \cite{Riess:2022mme} paper, the authors showed that there is very little correlation between the Cepheid metallicity-dependent correction to the PLR and the Hubble constant $H_0$. This robustness is primarily due to the uncertainties on $M_B$ and the inferred Hubble constant being dominantly influenced by the uncertainties on the corrected SNe apparent magnitudes and the limited statistics of the calibrators.

Thus, it is not surprising that our simplification of not including the metallicity correction does not significantly change the mean value or uncertainty of our inferred Hubble constant compared to \cite{Riess:2021jrx}. However, our approximation does lead us to underestimate the errors on the Cepheid PLR parameters and the fitted distance moduli.

Since one of our aims is to study whether a $G$-transition is a better fit to the distance ladder than the no $G$-transition hypothesis, we must be cautious of the fact that we have obtained a larger $\chi^2_{\textrm{dof}}$ than what we would have obtained if we had included the metallicity corrections. We will revisit this issue in Section~\ref{sec:comparison}, where we compare the quality of fits to both hypotheses.

In the next section, we discuss how we incorporated a $G$-transition hypothesis into our analysis, followed by a discussion of the $\chi^2$ fitting procedure for this alternate hypothesis.

\vspace{3mm}
\section{Effect of a $G$-transition on determination of Hubble Constant}\label{Gtrans-theory}

\subsection{Effect of a $G$-transition on the Cepheid PLR} 

If the value of the gravitational constant $G$ were different, it would impact both the pulsation period and the brightness of Cepheid variable stars. These changes would, in turn, affect the established relationship known as the Cepheid Period-Luminosity Relation (PLR).

The pulsation dynamics of Cepheids, which determine their pulsation period, are influenced by the Helium partial ionization zone in the star's envelope. On the other hand, the star's luminosity is governed by nuclear burning in its core~\cite{book_edi}. Therefore, we can analyze the changes in pulsation period and luminosity independently, as they are primarily driven by different processes.

Ritter \cite{ritter} demonstrated for the very first time that the pulsating period of a homogeneous sphere undergoing radial adiabatic pulsation is related to the mean surface density of the sphere as $P \propto \sqrt{R/g}$, where $g$ is the surface gravity and $R$ is the sphere's radius. Subsequent studies~\cite{book_edi, martin, edi18, edi19} extended this relationship to real stars. In simple terms, the pulsation period can be approximated to be proportional to the free-fall time of the Cepheid envelope, scaling as $P \propto 1/\sqrt{G \overline{\rho}}$~\cite{Sakstein2019}, where $\overline{\rho}$ is the mean density. Assuming that a change in $G$ does not alter the mean density, this leads to a scaling relation $P \propto 1/\sqrt{G}$. Consequently, if the effective $G$ changes by $\Delta G$, the change in Cepheid period is given by the equation:
\begin{equation} \label{eq:P_change}
\Delta \log (P) = - \frac{1}{2} \log \left(1 + \frac{\Delta G}{G_N}\right),
\end{equation}
where $G_N$ is the Newtonian gravitational constant. Specifically, for a positive change $\Delta G$, the period of the Cepheid would decrease.

Now, considering the change in luminosity due to a variation in $G$, Cepheid variables burn a H-shell surrounding an inert He core~\cite{book_edi}. For a fixed Cepheid mass, a slight increase in the effective gravitational constant would necessitate more pressure to maintain hydrostatic equilibrium. This additional pressure support can only be achieved through increased nuclear burning in the core, resulting in an overall increase in luminosity.

In their study, Sakstein et al.~\cite{Sakstein2019} conducted simulations utilizing the Modules for Experiments in Stellar Astrophysics (MESA) code, with modifications to the gravitational constant ($G$) in the cores of Cepheid stars. The MESA code, as described by Paxton et al.~\cite{Paxton_2019}, was employed for these simulations. The authors derived an expression characterizing the change in luminosity at the blue edge of the instability strip, represented by the equation:
\begin{equation} 
\label{eq:lum_change}
    \Delta \log L = B \log\left(1 + \frac{\Delta G}{G_N}\right),  
\end{equation}
where $\Delta \log L$ denotes the change in logarithmic luminosity, $\Delta G$ represents the modification in the gravitational constant, and $G_N$ is the nominal value of the gravitational constant. The coefficient $B$ in this equation is dependent on both the mass of the Cepheid star and the specific crossing of the instability strip being considered. In this context, "crossing" refers to the number of times the star traverses the instability strip in the Hertzsprung-Russell (HR) diagram. The typical range of values for the coefficient $B$ obtained by Sakstein et al. was found to be between $3.46$ and $4.52$. Notably, the positive nature of $B$ implies that an increase in the effective gravitational constant corresponds to an augmentation in luminosity.

\begin{figure*}
\vspace{-0.1cm}
\centering
\resizebox{270pt}{190pt}{\includegraphics{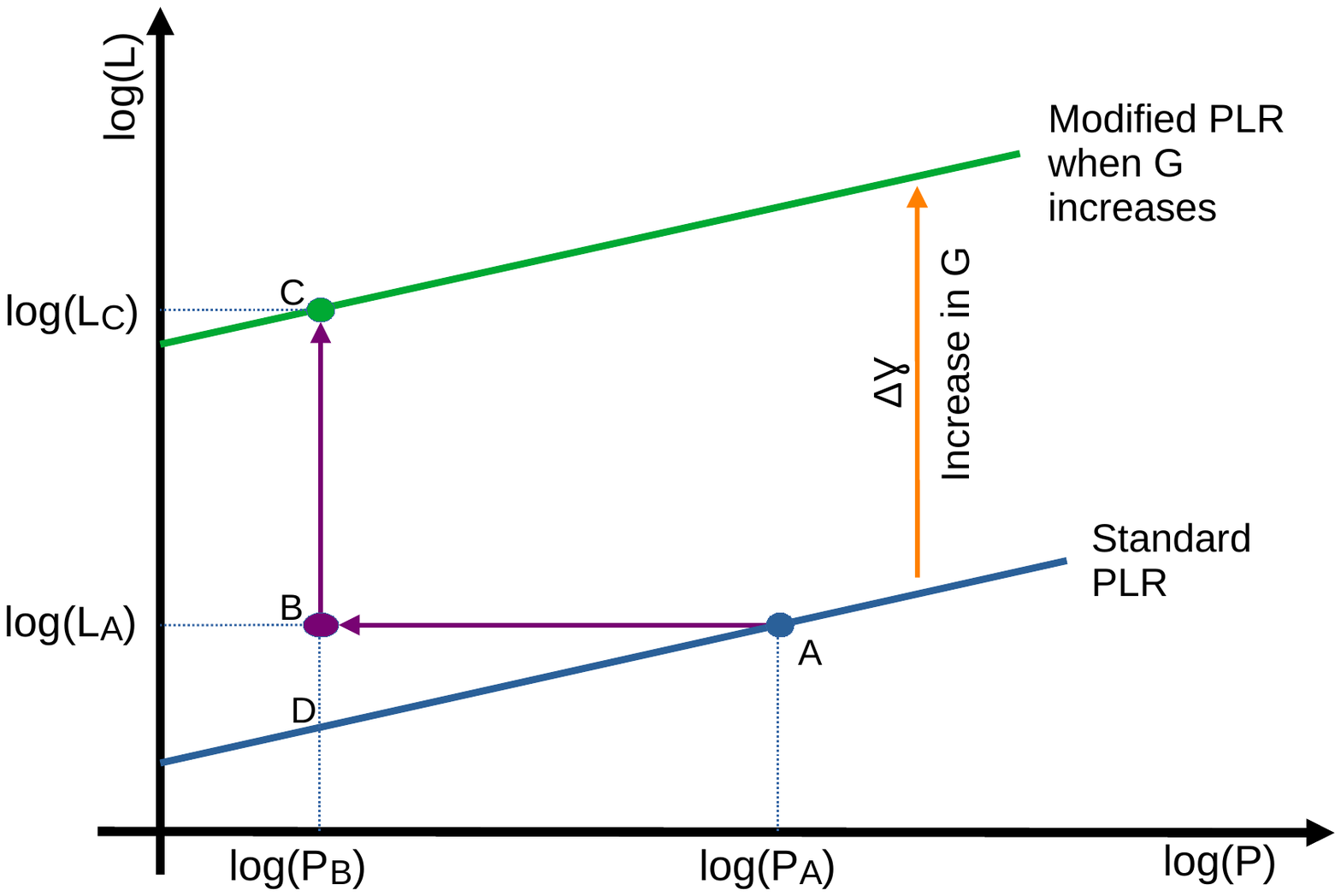}}
\caption{This illustration shows the change in the intercept of Cepheid PLR when standard $G$ is modified.}
\label{fig:PLRchange}
\end{figure*}

The influence of a variation in $G$ on the Period-Luminosity Relation (PLR) of Cepheid stars can now be comprehended by considering the combined effects on the period and luminosity of a given Cepheid star shown in Fig.~\ref{fig:PLRchange}. To understand the illustration in more depth, we refer you to check the schematic diagram of the Cepheid PLR (Fig.~1) in a recent similar study by Ruchika et al. \cite{Ruchika:2023ugh}. This plot shows what exactly happens to the luminosity and period for a Cepheid after a change in $G$. Conducting a systematic application of the outlined procedure to all Cepheids within the initial Period-Luminosity Relation (PLR), the resultant adjusted PLR is depicted as the green curve in the figure:
\begin{equation}
\label{eq:basic_PLR_with_G}
\log(L) = \alpha \log\left(\frac{P : \text{(days)}}{10 : \text{days}}\right) + \gamma + \Delta \gamma,
\end{equation}
where the alteration in the PLR intercept, denoted as $\Delta \gamma$, is expressed as:
\begin{equation}
\label{eq:PLR_intercept_G_1}
\Delta \gamma = \left(\frac{\alpha}{2} + B \right) \log \left (1 + \frac{\Delta G}{ G_N} \right) .
\end{equation}

Hence, the cumulative impact of a positive $(\Delta G > 0)$ transition in $G$ manifests an increased intercept in a revised Cepheid PLR while keeping the slope unaltered.

\vspace{3mm}

\subsection{Effect of a $G$-transition on the SNe~Ia standardized peak luminosity}

Modeling Type Ia Supernova (SNe Ia) explosions proves challenging due to the intricate nature of the explosion, including potential spontaneous transitions to detonation \cite{Pinto}. While numerical models could, in principle, be employed to investigate the impact of a gravitational constant ($G$) variation on the anticipated standardized peak luminosity of SNe Ia, we opt for a more straightforward approach by making simplified assumptions to derive an analytical expression for the change in standardized peak luminosity resulting from a $G$ transition.

An initial hypothesis regarding the dependence of SNe Ia standard luminosity on $G$ is to consider a direct proportionality to the Chandrasekhar mass $M_\textrm{Ch}$ \cite{woosley, amendola, Gazt}. This mass closely aligns with the Chandrasekhar limit $M_c \approx 1.44 M_\odot$ \cite{CO1996}, where relativistic degeneracy pressure becomes insufficient to prevent gravitational collapse. Notably, $M_c \sim G^{-3/2}$ \cite{CO1996}, and the inverse relationship of $M_c$ with $G$ is conceptually straightforward. A lower $G$ value results in a diminished gravitational pull per unit mass, enabling electron degeneracy pressure to counteract the gravitational pull from a larger mass just before collapse. Consequently, a higher mass star can resist gravitational collapse, leading to a higher $M_c$ for lower $G$. We assume that $M_\textrm{Ch}$ follows a similar scaling with $G$.

These assumptions suggest that the standardized SNe Ia luminosity $L$ scales as $G^{-3/2}$, indicating a decrease in luminosity with an increase in $G$. However, contrary to this expectation, a semi-analytic model of SNe light curves proposed by Wright and Li \cite{Wright2018}, as mentioned in the introduction, suggests that the standardized SNe Ia luminosity might actually increase for larger values of $G$. For a more detailed theoretical explanation of how it could happen, we again refer to Ruchika et al. \cite{Ruchika:2023ugh}.

Ref.~\cite{Sakstein2019} conducted a fitting analysis based on the outcomes of \cite{Wright2018}, establishing a scaling relation for the standardized luminosity of Type Ia Supernovae (SNe Ia) with respect to the gravitational constant $G$:
\begin{equation}
\label{eq:SNeIa_L_G}
    L \sim G^{1.46},
\end{equation}
This relation implies $L \propto M_c^{-0.97}$, indicating a decrease in standardized luminosity with the increased Chandrasekhar mass.

To accommodate the diverse scenarios discussed regarding the relationship between $L$ and $M_c$, we adopt a versatile approach, assuming $L \propto M_c^n \propto G^{-3n/2}$.

If the proposition of a $G$-transition at a distance $d_t$ holds true, it implies the existence of \textit{two distinct} standardized peak luminosities for SNe Ia. We denote $L_1$ as the standardized peak luminosity for SNe with $d < d_t$ and $L_2$ for those with $d > d_t$.

The disparity between these standardizations can be expressed as:
\begin{equation}
\label{eq:SNLcorr}
 (\textrm{log } L_2) - (\textrm{log } L_1) = -\frac{3 n}{2} \textrm{log} \left (1 + \frac{\Delta G}{G_{\textrm{N}}} \right).
\end{equation}

Considering that SNe Ia light curve standardization involves Hubble flow SNe at distances greater than 40~Mpc, and there is no observed evolution in the standardized light curve properties in the Hubble flow SNe, it implies that the $G$-transition occurs at distances $d_t$ less than $40$~Mpc.

For $n < 0$, as suggested in \cite{Wright2018}, we anticipate $L_2 > L_1$ for $\Delta G > 0$. Incorrectly assuming a uniform standardized peak luminosity at all distances would consequently result in an underestimation of SNe peak luminosity in the Hubble flow under such circumstances.

\subsection{Effect of a $G$-transition on the measured value of the Hubble Constant}
\label{sec:intu_effect_H0} 

In order to measure $H_0$, we need to calibrate the luminosity of type Ia SNe in the Hubble flow. The key assumptions of this calibration are i) that the Cepheid PLR is valid at all distances and ii) that there is only one true value of the standardized type Ia SNe peak luminosity. Both these assumptions are violated if there is a $G$-transition at a distance $d_t$ between 7 - 40 Mpc, which lies in the set of calibrator galaxies.

Let us assume that $\Delta G$ is positive and $d_t$ is between 7 - 40 Mpc. We will additionally assume that the peak SN luminosity $L$ scaling with $M_c$ has index $n < 0$. Given that we choose our transition distance to lie in the calibrator box, distances to the galaxies that are affected by the increase in $G$ would be underestimated. In standard practice, the inference of the Hubble constant would require averaging over all calibrators which would correspond to one value of the standardized type Ia SNe peak luminosity. Since in our case, some of the galaxies would be affected by $G$-transition and the others will not be affected, we will now have two different values of standardized type Ia SNe peak luminosity averaged separately for the affected and unaffected galaxies.

In the next section, we will allow for a positive $G$-transition and check if it is preferred by the observational data.

\section{Fitting the distance ladder to a $G$-transition}
\label{sec:Gtransitionfit}

We now discuss the hypothesis of a $G$-transition at a lookback time corresponding to a luminosity distance $d_t$ which lies between $7$ and $40$~Mpc, or equivalently a distance modulus $\mu_t$ which lies between 29 and 33\footnote{Although the calibrator step extends from $\mu = 28.5$ to $\mu = 33$, we do not consider values of $\mu_t$ at the extreme ends of the calibrator step, i.e., $\mu_t = 28.5$ or 33. This is because if we take $\mu_t = 28.5$, all the calibrator galaxies will be affected by the $G$-transition. Alternatively, if we take $\mu_t = 33$, there will be no calibrator galaxy that is affected by the $G$-transition.}. This distance range corresponds to a transition in the Cepheid calibrator step of the distance ladder. We will assume that $G$ was larger than $G_N$ in the past by an amount $\Delta G$. 

In order to completely specify the hypothesis of $G$-transition, we need to specify the values of $\mu_t$ and $\frac{\Delta G}{G_N}$. We will consider several such hypotheses corresponding to a different choice of these parameters, where we select $\mu_t$ from a list of values between 29 and 32.5 in steps of 0.5, and positive fractions $\frac{\Delta G}{G_N}$ from the following possibilities 2\%, 4\%, 5\%, 6\%, and 8\%. Thus, in total we are considering 40 different combinations of these two parameters. Each one of these points represents a fully specified hypothesis for a $G$-transition.

The main change in our analysis with a $G$-transition is that our predictions of the Cepheid and SNe apparent magnitudes have to be altered.

The Cepheids in the anchors, and in calibrator galaxies residing at a $\mu < \mu_t$ are not expected to be affected by the $G$-transition, and we can use eqs.~\ref{eq: N4258_PLR_model}, \ref{eq: LMC_PLR_model}, \ref{eq: MW_PLR_model}, for the anchor Cepheids, and eq.~\ref{eq: plr_model} for the Cepheids in the calibrators. However, for Cepheids in the calibrator galaxies residing at $\mu > \mu_t$, the PLR intercept will get modified because of the effect of the $G$-transition (eq.~\ref{eq:basic_PLR_with_G}). For these Cepheids, the prediction of their apparent magnitudes from the PLR model is given by,
\begin{equation}
\label{eq:PLR_intercept_G_2}
m_{H,i,j}^W = 
\mu_{i} +  M^W_{H,1} + \left[\frac{b_W}{2}\log \left (1 + \frac{\Delta G}{G_N} \right ) -2.5B\log \left (1 + \frac{\Delta G}{G_N} \right)\right] + b_W (\log{(P_{i,j})} - 1).
\end{equation}
Here, the coefficient $B$ depends on the Cepheid mass. It also depends on whether the Cepheid is observed at the second or third crossing of the instability strip. Sakstein et al.~\cite{Sakstein2019} have tabulated values of $B$ as a function of the stellar mass and the instability strip crossing epoch. The values of $B$ range from $3.46$ to $4.52$. In our analysis, we take an intermediate value $B = 4$. Though later in this work, we also mention how our results may change when we take the extreme values of $B$.

Similarly for SNe in the calibrators at $\mu < \mu_t$, we can continue to use eq.~\ref{eq:SNeIaMB}, with $M_B$ replaced by $M_{B1}$ to denote that this standardized absolute magnitude is valid only up till $\mu_t$. For SNe in calibrators with $\mu > \mu_t$, we can again use eq.~\ref{eq:SNeIaMB}, but with $M_B$ replaced by $M_{B2}$, which denotes the standardized absolute magnitude for distant SNe (for calibrators as well as Hubble flow supernovae).

Note that the distance moduli to the calibrator galaxies are undetermined a priori, so the appropriate case for these formulae (to the left or the right of the transition) has to be used when scanning over the values of these distance parameters in an MCMC analysis.

Our new $\chi^2$ study which incorporates a $G$-transition has the following fit parameter dependence: $b_W$ (slope of PLR), $M_W^H$ (intercept of PLR), $\mu_i$ (distance moduli for the 19 calibrator galaxies hosting Cepheids and Type Ia SNe), $M_{B1}$, and $M_{B2}$.

The Hubble constant estimate can be deduced from our fit using Eq.~\ref{eq:H0MB} with $M_B$, now replaced by $M_{B2}$, specifically employing the standardized peak luminosity for distant supernovae (situated to the right of $\mu_t$).

We refrain from imposing a fixed relationship between $M_{B1}$ and $M_{B2}$ during the fitting process, opting instead to treat them as free parameters. This approach is akin to allowing the index $n$ of the Type Ia supernovae (SNe Ia) $L-M_c$ relation to be determined as a derived parameter from the fit. Once we obtain the constraints on $M_{B1}$ and $M_{B2}$, the value of $n$ can be easily determined by inversely applying Eq.~\ref{eq:SNLcorr} as:

\begin{equation}
\label{eq:nvalue}
n = \frac{2}{7.5} \frac{M_{B2}-M_{B1}}{\log\left(1 + \frac{\Delta G}{G_N}\right)}.
\end{equation}

\subsection{Parameter priors}
\label{sec:parameter_priors_Gtransition}

For a given value of $\Delta G$ and $\mu_t$, we are fitting exactly the same parameters as before in the case of the no $G$-transition hypothesis, with the exception of $M_B$. $M_B$ is now replaced by the parameters $M_{B1}$ and $M_{B2}$, for each of these we use a uniform prior ranging from $-21$ to $-17$~mag. For other distance ladder parameters, we use the same priors as in Table~\ref{table:basic_prior}.

\subsection{$\chi^2$ minimisation}
For each hypothesis of $\Delta G$ and $\mu_t$ we can now fit our distance ladder to obtain the fit parameters by minimizing a $\chi^2$. We define our $\chi^2$ exactly the same way as in the case of the no $G$-transition hypothesis (see eqs.~\ref{eq:Lcep}, \ref{eq:LSN}, and \ref{eq:Ltotal}). The only difference now is that for each term in the $\chi^2$, we use the theoretical predictions of the modified expressions for the Cepheid and SNe apparent magnitudes, where the modification is due to the $G$-transition effect.

Note that we have 25 parameters that we are minimizing over -- the increase of one parameter compared to the case of no $G$-transition comes from the fact that we are using two parameters $M_{B1}$ and $M_{B2}$ for the standardized SNe peak luminosities to the left and to the right of the transition.

Note that $\Delta G$ and $\mu_t$ are not parameters that we are fitting for, since we are considering each combination of these to be separate hypotheses. However, when comparing the quality of fits of these models to the data with the fit to the distance ladder without a $G$-transition, we will appropriately penalize for the ``look-elsewhere'' effect of scanning over a grid of such parameters.

\subsection{Results for the distance ladder fit in the presence of a $G$-transition}
\label{sec:results}

\begin{table*}
\centering
\scalebox{0.7}{
\renewcommand{\arraystretch}{2.3}
\begin{tabular}{|p{0.8cm} |p{0.7cm}||p{1.9cm}|p{1.3cm}||p{1.9cm}|p{1.3cm} ||p{1.9cm}|p{1.3cm} || p{1.9cm}|p{1.3cm} ||p{1.9cm}|p{1.3cm} ||}
 \hline
  \multicolumn{2}{|c||}{ } &   \multicolumn{2}{|c||}{ $\Delta G/G_{\textrm{N}} = 2 \%$}&  \multicolumn{2}{|c||}{ $\Delta G/G_{\textrm{N}} = 4 \%$}&   \multicolumn{2}{|c||}{ $\Delta G/G_{\textrm{N}} = 5 \%$}&   \multicolumn{2}{|c||}{ $\Delta G/G_{\textrm{N}} = 6 \%$}&\multicolumn{2}{|c||}{ $\Delta G/G_{\textrm{N}} = 8 \%$}\\
 \hline
 $\mu_t$& & &$\chi^2_\textrm{min}$& &$\chi^2_\textrm{min}$& &$\chi^2_\textrm{min}$ & &$\chi^2_\textrm{min}$& &$\chi^2_\textrm{min}$ \\
 \hline
 \hline
   $29$ & $H_0$ &$70.33\pm1.06$ & $1963.92$&$67.53 \pm 1.06$ & $1963.72$&$66.02\pm0.98$&$1963.59$& $64.54\pm 0.96$ & $1963.67$& $61.81 \pm0.82$ & $1963.68$ \\ \cline{2-3} \cline{5-5}\cline{7-7}\cline{9-9}\cline{11-11}
  $ $&$n$ &$-3.41\pm5.81$ & &$-3.09\pm3.30$& & $-2.95\pm1.54$& &$-2.96\pm1.80$ & & $-3.15\pm1.04$&  \\
 \hline
 \hline
  $29.5$ & $H_0$ &$70.76\pm1.18$ & $1963.23$&$67.91\pm0.89$ & $1963.21$&$66.14\pm1.10$&$1963.24$& $64.81\pm0.82$ & $1963.18$& $61.91\pm0.98$ & $1963.50$ \\ \cline{2-3} \cline{5-5}\cline{7-7}\cline{9-9}\cline{11-11}
    $ $&$n$ &$-0.51\pm2.79$ & &$-2.17\pm1.37$& & $-2.01\pm1.46$& &$-2.01\pm1.11$ & & $-2.50\pm0.81$&  \\
 \hline
 \hline
   $30$ & $H_0$ &$70.63\pm1.03$ & $1962.92$&$67.62\pm1.02$ & $1963.18$&$66.16\pm1.14$&$1963.06$& $64.86\pm0.99$ & $1963.39$& $62.03\pm0.85$ & $1963.22  $ \\ \cline{2-3} \cline{5-5}\cline{7-7}\cline{9-9}\cline{11-11}
  $ $&$n$ &$-0.52\pm2.84$ & & $-1.69\pm1.58$ & &$-1.87\pm1.56$& &$-2.30\pm1.02$& & $-2.38\pm0.90$&\\
 \hline
 \hline  
 
   $30.5$ & $H_0$ &$70.66\pm1.17$ & $1963.20$&$67.72\pm1.1$ & $1963.02$&$66.57\pm0.89$&$1963.58$& $64.87\pm0.94$&$1963.28$& $61.97\pm0.95$ & $1963.32$ \\ \cline{2-3} \cline{5-5}\cline{7-7}\cline{9-9}\cline{11-11}
  $ $&$n$ &$-0.32\pm2.97$ & &$-1.68\pm1.15$& &$-1.73\pm1.39$& & $-2.08\pm1.30$& & $-2.37\pm1.02$&\\
 \hline
 \hline   
 
   $31$ & $H_0$ &$70.79\pm1.21$ & $1962.65$&$67.81\pm1.03$ & $1963.07$&$66.47\pm1.01$&$1962.91$& $64.92\pm1.13$ & $1962.96$& $62.09\pm1.06$ & $1962.56$ \\ \cline{2-3} \cline{5-5}\cline{7-7}\cline{9-9}\cline{11-11}
  $ $&$n$ &$-0.20\pm2.23$ & &$-1.82\pm1.31$& &$-1.92\pm1.01$& & $-2.10\pm0.79$& & $-2.50\pm0.58$&\\
 \hline
 \hline

   $31.5$ & $H_0$ &$70.49\pm1.34$ & $1966.03$&$67.40\pm1.12$ & $1964.29$&$65.83 \pm 1.21$&$1964.18$& $64.66\pm1.29$&$1964.19$& $61.73\pm1.06$ & $1964.95$ \\ \cline{2-3} \cline{5-5}\cline{7-7}\cline{9-9}\cline{11-11}
  $ $&$n$ &$-3.20\pm2.20$ & &$-2.97\pm0.95$& &$-3.40 \pm 0.71$& & $-3.21\pm0.75$& & $-3.21 \pm 0.56$&\\
 \hline
 \hline

   $32$ & $H_0$ &$68.96\pm1.72$ & $1962.80$&\textbf{65.90}$\pm$\textbf{1.70} & \textbf{1962.13}&$64.60\pm1.68$&$1962.37$& $63.33\pm1.41$ & $1962.56$& $60.40\pm1.42$ & $1962.42$ \\ \cline{2-3} \cline{5-5}\cline{7-7}\cline{9-9}\cline{11-11}
  $ $&$n$ &$-5.36\pm2.00$ & &\textbf{-4.24}$\pm$\textbf{1.06}& &$-4.03\pm0.76$& & $-3.85\pm0.65$& & $-3.74\pm0.50$&\\
 \hline
 \hline  
 
   $32.5$ & $H_0$ &$68.87\pm2.51$ & $1964.32$&$65.86\pm2.02$ & $1964.28$&$64.56\pm1.95$&$1963.33$& $63.02\pm1.44$ & $1963.74$& $60.44\pm1.78$ & $1962.94$ \\ \cline{2-3} \cline{5-5}\cline{7-7}\cline{9-9}\cline{11-11}
  $ $&$n$ &$-4.93\pm2.52$ & &$-4.15\pm1.11$& &$-3.77\pm0.94$& & $-3.86\pm0.55$& & $-3.63\pm0.66$&\\
 \hline
 \hline 
  
\end{tabular}
}
\caption{This table shows a grid of different parameter choices for the hypothesis of a $G$-transition with transition distance modulus $\mu_t$ ranging from 29 to 32.5 and a positive $\Delta G /G$ varying between 2 to 8 percent. For each grid point, we performed a fit to the distance ladder, computed the $\chi^2_{\textrm{min}}$, and derived constraints on the indirectly inferred parameters $H_0$ (in units of km/s/Mpc) and $n$. The entries in the table show these values for each grid point. The entries highlighted in bold correspond to our benchmark point with $\mu_t = 32$ and $\Delta G/G_{\textrm{N}} = 4\%$, which has one of the lowest values of $\chi^2_{\textrm{min}}$. }\label{table:grid_results}
\end{table*}

We now present our results for our analysis of the 40 different combinations of the model parameters $\frac{\Delta G}{G_N}$ and $\mu_t$ of the $G$-transition hypothesis. For each combination, we minimize the total $\chi^2$ to obtain fits to the 25 fit parameters - 19 distances to calibrator galaxies, 2 distances to the LMC and NGC4258, the slope and intercept of the Cepheid PLR, and the SNe standardized luminosities to the left and the right of the transition, $M_{B1}$, $M_{B2}$. Once we obtain the values of $M_{B1}$ and $M_{B2}$, we can obtain the Hubble constant $H_0$ and the SN $L-M_c$ index $n$ as derived parameters.

In Table \ref{table:grid_results}, we show a snapshot of the results for all 40 model hypothesis combinations of $\frac{\Delta G}{G_N}$ and $\mu_t$. For each value of $\frac{\Delta G}{G_N}$ and $\mu_t$, we show the values of the minimum $\chi^2$, along with the $H_0$ and $n$ inferred from the best-fit parameters.

\begin{figure*}
\resizebox{440pt}{180pt}{\includegraphics{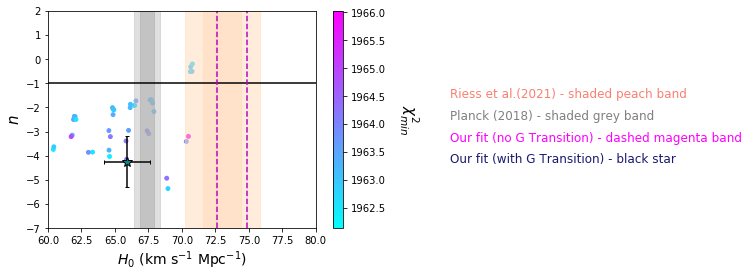}}
\vspace{.1cm}
\caption{This scatter plot is a pictorial representation of Table \ref{table:grid_results}. Here, we show the inferred values of the Hubble constant $H_0$ and the parameter $n$ which gives the index of the standardized SNe~Ia $L-M_c$ scaling relation $L \propto M_c^n$. Each point in the scatter plot corresponds to a particular grid point in the $\mu_t-\Delta G/G_{\textrm{N}}$ grid. The colors of the points indicate the minimum $\chi^2$ associated with the corresponding grid point. The dashed magenta lines show for comparison our constraints on $H_0$ corresponding to our fit when we do not assume a $G$-transition, for which $\chi^2_{\textrm{min}}$ is $1963.19$ as discussed in Section~\ref{sec:resultsnoG}. The peach-colored band shows the 1$\sigma$ and 2$\sigma$ regions of $H_0$ from the \cite{Riess:2021jrx} fit, and our results in the case of no $G$-transition are in good agreement with theirs. The grey bands show the 1$\sigma$ and 2$\sigma$ regions of $H_0$ from the Planck 2018 results \cite{Planck:2018vyg} respectively. Our benchmark point with a gravitational transition at $\mu_t = 32$ and $\Delta G/G_{\textrm{N}} = 4\%$ is shown with a black dot with error bars indicating the uncertainty on the best-fit inference of $H_0$ and $n$. This point has one of the lowest $\chi^2$ values ($1962.13$) in our entire grid scan. This benchmark point also has a value of $H_0$ that is in good agreement with the Planck 2018 results, and a value of $n$ that is within 2-$\sigma$ agreement with the value $n = -0.973$ (solid black horizontal line) obtained in the study of \cite{Wright2018}.}
\label{fig:5dfigure}
\end{figure*}

To easily visualize the data in this table, we have represented the $G$-transition model parameter points in a scatter plot of the inferred $H_0$ and $n$ values that they yield for the best-fit parameters of each in Fig.~\ref{fig:5dfigure}. The points are color coded to encode the value of $\chi^2_{\textrm{min}}$. We have not shown the error bars on $H_0$ and $n$ for each point to avoid cluttering the plot.

For comparison, we also show in the same figure, the values of $H_0$ inferred using our fit to the no $G$-transition hypothesis, which agrees with that of \cite{Riess:2021jrx} (also shown). The $H_0$ value inferred from observations of the CMB by the Planck data~\cite{Planck:2018vyg} is also shown as a grey band. Among the model parameter points with the lowest $\chi^2_\textrm{min}$, the point with $\mu_t = 32$, and $\frac{\Delta G}{G_N} = 4\%$ is particularly intriguing. This model point is indicated in bold font in Table~\ref{table:grid_results} and with a black dot (with error bars shown) in Fig.~\ref{fig:5dfigure}. It yields a value of $H_0 = 65.90 \pm 1.70$~km/s/Mpc, which is in good agreement with the value obtained from the CMB. For this point, the value of $n$ is found to be $-4.24 \pm 1.06$, which has the same sign as the semi-analytic prediction of Ref.~\cite{Wright2018} of $n=-0.97$. Furthermore, as we shall discuss later, the best-fit point of this model is consistent with the constraint $\frac{\Delta G}{G_N} < 5\%$ obtained from CMB and BBN bounds (see sec.~\ref{sec:discussion}). Given these observations, we will analyze this $G$-transition parameter point further in-depth, and we will refer to it as our favored $G$-transition model.

\begin{figure*}
\vspace{0.1cm}
\resizebox{220pt}{180pt}{\includegraphics{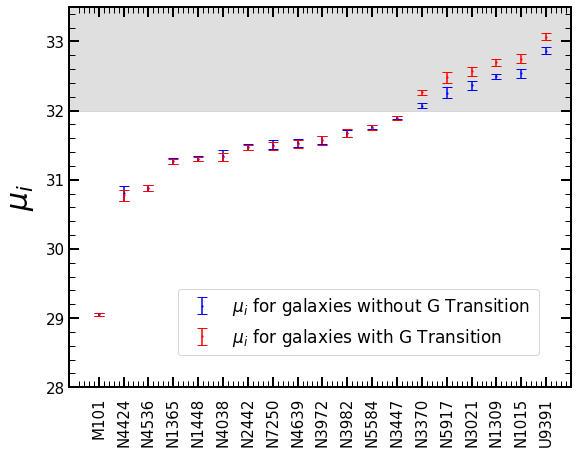}}
\vspace{0.1cm}
\resizebox{220pt}{180pt}{\includegraphics{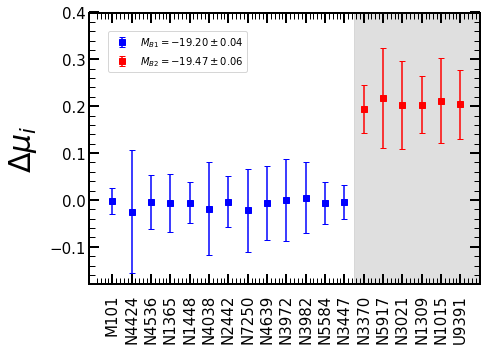}}
\caption{The left panel shows our best-fit distance moduli (with uncertainties) to various galaxies in the calibrator sample in the case without a $G$-transition and with a $4\%$ increase in $G$ at $\mu_t = 32$. The right panel shows the difference between the magnitude of the distance moduli of galaxies of the left plot between the cases with and without a $G$-transition. The shaded region in both the plots shows distances greater than the transition distance, i.e., $\mu > \mu_t$.}
\label{fig:mu_compare_both}
\end{figure*}

For the point with $\mu_t = 32$ and $\frac{\Delta G}{G_N} = 4\%$, we present the best-fit values of the main fit parameters in Table~\ref{table:results_table_final}. This table does not show the best-fit values of the distance moduli to the 19 calibrator galaxies, which we show instead in Table~\ref{table:result_G_mu} and Fig.~\ref{fig:mu_compare_both}.

To compare our best-fit parameters for the no $G$-transition hypothesis, we also show the corresponding values (where applicable) from our analysis in Section~\ref{sec:resultsnoG} in Table~\ref{table:results_table_final} and Fig.~\ref{fig:mu_compare_both}. We also compute the differences in distance moduli to the calibrators between the two hypotheses, along with the error (calculated by adding the errors in each model in quadrature) in the right panel of Fig.~\ref{fig:mu_compare_both}.

In both panels of Fig.~\ref{fig:mu_compare_both}, the grey shaded region represents the calibrators whose inferred distance would be affected by a $G$-transition at $\mu_t = 32$. We can clearly see from the figure that calibrator galaxies to the right of the transition would be incorrectly inferred to have a shorter distance (under the no $G$-transition fit) than their true distance if the hypothesis of this $G$-transition is true. This difference in the inferred distances to calibrators between the fits with and without a $G$-transition leads to a lower inferred value of $M_{B2}$ compared to $M_B$, and hence to a smaller value of $H_0$ in the case of a $G$-transition.

\begin{figure*}
\vspace{-0.1cm}
\resizebox{450pt}{490pt}{\includegraphics{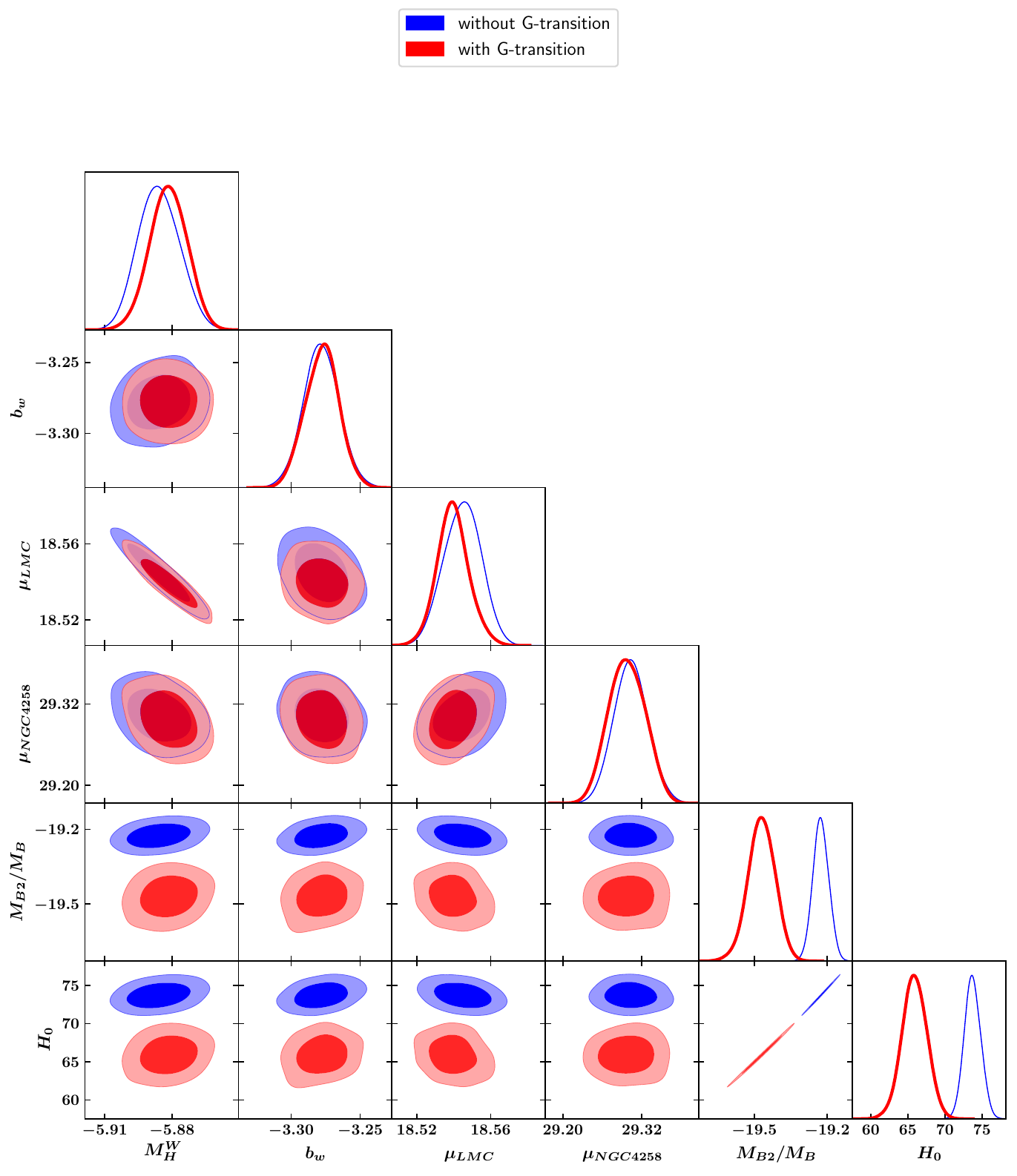}}
\caption{This corner plot shows the correlations between various model fit parameters along with their 1D posteriors, in both the cases with (red) and without (blue) a $G$-transition. The $G$-transition is taken to occur at $\mu_t = 32$ with $\Delta G/G_{\textrm{N}} = 4\%$. We have also shown the correlations with the derived parameter $H_0$ which is inferred from $M_B/M_{B2}$. For the case with no $G$-transition, the SNe~Ia standardized absolute magnitude is represented by $M_B$, whereas in the case of a $G$-transition, there are two values of absolute magnitudes to the left and to the right of the transition: $M_{B1}$ and $M_{B2}$ respectively. $M_{B2}$ is what is used to infer the value of $H_0$ because it corresponds to the absolute magnitude of the Hubble flow SNe. Hence, we have shown $M_{B2}$ in the figure for the case with a $G$-transition instead of $M_B$.}
\label{fig:corner_plot}
\end{figure*}

We have also shown a corner plot to illustrate the correlation between the most important fit parameters in Fig.~\ref{fig:corner_plot} in red. To compare the values and correlations with those of the no $G$-transition hypothesis, we have also shown the correlations from our fit in Section~\ref{sec:resultsnoG} in blue. The parameters shown in the plot have a one-to-one correspondence with each other in each of the respective fits. For the no $G$-transition case, there is only one standardized SN luminosity $M_B$, which we show in the corner plot. However, for the $G$-transition fit, there are two parameters, $M_{B1}$ and $M_{B2}$. We correspond $M_B$ of the no $G$-transition fit with the parameter $M_{B2}$, which represents the standardized SN peak luminosity to the right of the $G$-transition, since this is the value used for the Hubble flow SNe. Although $H_0$ is a derived parameter in the fits, exactly inferred from $M_B/M_{B2}$, we have shown its posterior distribution in the same plot, as this is the parameter of primary interest.

From the plot, and also from Table~\ref{table:results_table_final}, we can see that there is no significant change (within 1-$\sigma$) in the parameters for distances to the LMC and NGC4258, as well as in the Cepheid PLR parameters. The main change between the fits is the difference between the standardized Hubble flow SNe luminosities $M_{B2}$ and $M_B$. As anticipated, since $M_{B2}$ is inferred to be smaller than $M_B$, Hubble flow SNe are inferred to be intrinsically brighter in the $G$-transition model and are hence inferred to be further away compared to the no $G$-transition scenario. This leads to a correspondingly smaller inferred Hubble constant in the case of the $G$-transition hypothesis.

\begin{figure*}
\vspace{0.1cm}
\resizebox{220pt}{180pt}{\includegraphics{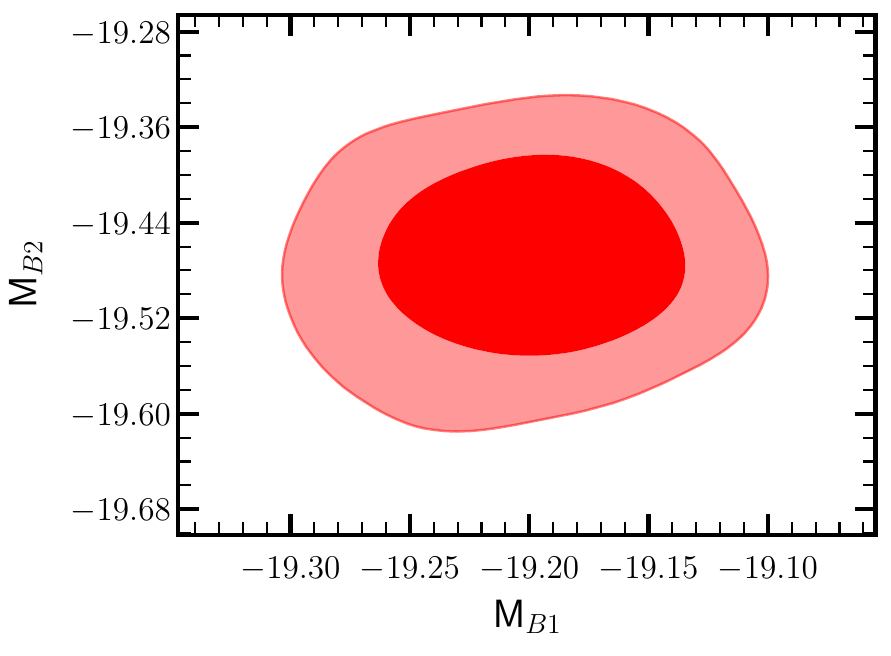}}
\vspace{0.1cm}
\resizebox{220pt}{180pt}{\includegraphics{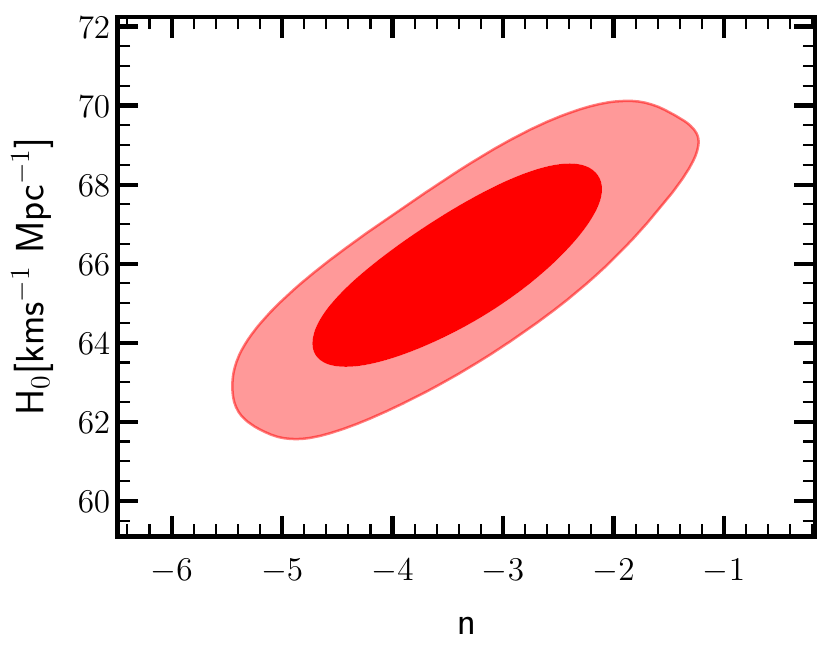}}
\caption{This figure shows the 2D correlations between the fit parameters $M_{B1}$-$M_{B2}$ (left panel) and the derived fit parameters $H_0$-$n$ (right panel) when assuming the $G$-transition hypothesis with $\mu_t = 32$ and $\Delta G/G_{\textrm{N}} = 4\%$.}
\label{fig:3d_plot}
\end{figure*}

In the corner plot of Fig.~\ref{fig:corner_plot}, we have not shown the parameter $M_{B1}$, which is unique to the $G$-transition hypothesis. The value of $M_{B1}$ for this set of model parameters ($\Delta G = 4\%$ and $\mu_t = 32$) is determined primarily from the SNe in the 13 calibrator galaxies to the left of the $G$-transition (see Fig.~\ref{fig:mu_compare_both}), whereas $M_{B2}$ is determined by the SNe in the 6 calibrator galaxies to the right of the transition. Thus, we do not expect any significant correlation between $M_{B1}$ and $M_{B2}$. To confirm this low correlation visually, we have plotted the 2-D posterior correlation between the parameters $M_{B1}$ and $M_{B2}$ in Fig.~\ref{fig:3d_plot} (left panel).

The value of the supernova $L-M_c$ scaling index $n$ is a derived parameter of our fits, which can be obtained from Eq.~\ref{eq:nvalue}. We find an inferred SN scaling relation $L \propto M_c^{-4.24 \pm 1.06}$. In Fig.~\ref{fig:3d_plot} (right panel), we show a correlation plot between $H_{0}$ and $n$, along with the 1-$\sigma$ and 2-$sigma$ contours for each. Since $H_0$ is determined from $M_{B2}$ and $n$ depends on both $M_{B1}$ and $M_{B2}$, we expect some correlation between $H_0$ and $n$.

\begin{table}
\centering
\scalebox{0.9}{
\renewcommand{\arraystretch}{2.3}
\begin{tabular}{|c|c|c|}
\hline
\hline
Parameters & Without $G$-transition & With $G$-transition \\ 
\hline
\hline
$M^W_H$ & $-5.89 \pm 0.01$ & $-5.88 \pm 0.01$ \\ \hline
$b_W$  & $-3.27 \pm 0.01$ & $-3.28 \pm 0.01$ \\ \hline
$\mu_{LMC}$  & $18.54 \pm 0.01$ & $18.54 \pm 0.01$ \\ \hline
$\mu_{NGC4258}$ & $29.30 \pm 0.03$ & $29.30 \pm 0.03$ \\ \hline
$M_{B}$ & $-19.23 \pm 0.03$ & $-$ \\ \hline
$M_{B1}$ & $-$ & $-19.20 \pm 0.04$ \\ \hline
$M_{B2}$ & $-$ & $-19.47 \pm 0.06$ \\ \hline
$n$ & $-$ & $-4.24 \pm 1.06$ \\ \hline
$H_0$ (km/s/Mpc) & $73.73 \pm 1.12$ & $65.90 \pm 1.70$ \\ \hline
\hline
\end{tabular}}
\caption{This table shows the comparison between best-fit values and 1-$\sigma$ intervals on various fit/inferred parameters for the distance ladder for both the cases with and without a $G$-transition. In the case of the $G$-transition, we are selecting our benchmark point with $\mu_t = 32.0$ and $\Delta G/G_{\textrm{N}} = 4\%$.}
\label{table:results_table_final}
\end{table}

\begin{table}
    \centering
\scalebox{0.85}{
\renewcommand{\arraystretch}{1.5}
\begin{tabular}{|c|c|c|c|c|c|c|c|c|}
\hline
\hline
S. No. & Galaxy & Without $G$-transition & With $G$-transition & S. No. & Galaxy & Without $G$-transition & With $G$-transition \\ 
\hline
\hline
1 & M101 & $29.05 \pm 0.02$ & $29.05 \pm 0.021$ & 11 & N3982 & $31.67 \pm 0.05$ & $31.68 \pm 0.053$ \\ 
2 & N4424 & $30.80 \pm 0.10$ & $30.78 \pm 0.081$ & 12 & N5584 & $31.76 \pm 0.03$ & $31.75 \pm 0.030$ \\ 
3 & N4536 & $30.88 \pm 0.04$ & $30.87 \pm 0.041$ & 13 & N3447 & $31.90 \pm 0.03$ & $31.89 \pm 0.024$ \\ 
4 & N1365 & $31.27 \pm 0.05$ & $31.26 \pm 0.042$ & 14 & N3370 & $32.07 \pm 0.04$ & $32.26 \pm 0.036$ \\ 
5 & N1448 & $31.30 \pm 0.03$ & $31.30 \pm 0.028$ & 15 & N5917 & $32.26 \pm 0.08$ & $32.48 \pm 0.074$ \\ 
6 & N4038 & $31.35 \pm 0.08$ & $31.33 \pm 0.060$ & 16 & N3021 & $32.37 \pm 0.06$ & $32.57 \pm 0.067$ \\ 
7 & N2442 & $31.47 \pm 0.04$ & $31.47 \pm 0.036$ & 17 & N1309 & $32.49 \pm 0.04$ & $32.69 \pm 0.046$ \\ 
8 & N7250 & $31.51 \pm 0.06$ & $31.49 \pm 0.062$ & 18 & N1015 & $32.54 \pm 0.06$ & $32.75 \pm 0.066$ \\ 
9 & N4639 & $31.53 \pm 0.05$ & $31.52 \pm 0.058$ & 19 & U9391 & $32.86 \pm 0.05$ & $33.07 \pm 0.054$ \\ 
10 & N3972 & $31.57 \pm 0.06$ & $31.57 \pm 0.065$ & & & & \\ 
\hline
\end{tabular}}
\caption{Comparison between the best-fit value and 1-$\sigma$ error on distance moduli to the 19 calibrator galaxies obtained from our fit without and with $G$-transition. Our analysis has tried to emulate that of \cite{Riess:2021jrx}, and we find good consistency despite making various simplifications in the formulation of the standard distance ladder. These simplifications lead to minor differences between our parameter estimates and those of the SH0ES team.}
\label{table:result_G_mu}
\end{table}

\section{Model Comparison}
\label{sec:comparison}

In the preceding section, our investigation highlighted the $G$-transition model with $\Delta G = 4\%$ and $\mu_t = 32$ as providing the best-fit to Cepheids and SNe Type Ia datasets compared to alternative parameter combinations within the $G$-transition framework. Remarkably, this specific parameter configuration also yielded the most fitting value for the Hubble constant, $H_0$, aligning with Cosmic Microwave Background (CMB) inferences of this cosmological parameter. Nonetheless, to assert its potential as a resolution to the Hubble tension, a pivotal inquiry arises: Does our favored $G$-transition model, characterized by $\Delta G = 4\%$ and $\mu_t = 32$, present a superior fit to the data when contrasted with the hypothesis of no $G$-transition?

This critical question is addressed through the application of three distinct estimators gauging the quality of fit, and a comparative analysis of their outcomes between the two hypotheses. The definitions of these estimators are as follows:

\begin{eqnarray}
\chi^2_{\textrm{dof}} &=& \frac{1}{N-d} \chi^{2}_{\textrm{min}}, \nonumber \\
\textrm{AIC} &=& \chi^{2}_{\textrm{min}} + 2d, \nonumber\\
\textrm{BIC} &=& \chi^{2}_{\textrm{min}} + d \ln(N).
\end{eqnarray}

The primary estimator utilized is the $\chi^2$ per degree of freedom, denoted as $\chi^2_{\textrm{dof}}$. The other two employed are the well-established Akaike Information Criterion (AIC)~\cite{Akaike} and Bayesian Information Criterion (BIC)~\cite{Schwarz:1978tp}. To articulate these estimators, we necessitate the minimum chi-squared value $\chi^2_\textrm{min}$, the count of model parameters $d$, and the number of data points $N$. The $G$-transition hypothesis introduces two standardized luminosity parameters for supernovae, namely $M_{B1}$ and $M_{B2}$, in contrast to the single parameter $M_B$ characterizing the no $G$-transition scenario. Moreover, while specifying the values of the $G$-transition model parameters $\Delta G$ and $\mu_t$, we have explored various plausible combinations for these values. Consequently, our estimators must account for the presence of 3 additional parameters associated with the $G$-transition model.

The model with the $\chi^2_{\textrm{dof}}$ value closer to one is preferred by definition. For AIC or BIC, the criteria are defined by the Jeffreys scale. In Table \ref{table:AIC_BIC}, we have shown the values of each of these estimators for the model without a $G$-transition and the model with a $G$-transition at $\Delta G/G_{\textrm{N}} = 4\%$ and $\mu_t = 32$. We find nearly the same $\chi^2_{\textrm{dof}}$ for the standard distance ladder and the one with the $G$-transition hypothesis. However, AIC and BIC still prefer the fit using the standard distance ladder in a standard gravity scenario.

\begin{table*}
\centering
\begin{tabular}{cccccccccc}
\hline \hline
\vspace{2mm}
Model & $\chi^2_{\rm min}$ & $d$ & $N$ & $\chi^2_{d.o.f}$ & ${\rm AIC}$ & ${\rm BIC}$ & $\Delta {\rm AIC}$ & $\Delta {\rm BIC}$ \\
\hline
Basic & $1963.19$ & $24$ & $1267$ & $1.579$ & $2011.19$ & $2134.65$ & $0.0$ & $0.0$ \\
\hline
With $G$-transition & $1962.13$ & $27$ & $1267$ & $1.582$ & $2016.13$ & $2155.02$ & $4.94$ & $20.37$ \\
\hline \hline
\end{tabular}
\caption{The values of $\chi^2_{\textrm{dof}}$, AIC, and BIC obtained for the distance ladder model with and without a $G$-transition. Lower values of each parameter indicate a better fit to the data. The results with a $G$-transition are shown for $\mu_t=32$ and $\Delta G/G_{\textrm{N}} = 4\%$.}
\label{table:AIC_BIC}
\end{table*}

\section{Summary, discussion and future studies}
\label{sec:discussion}

In this work, we questioned how standard the standard distance ladder is. What are the components of the Distance Ladder that we can blindly assume as standard? What happens if we try to change them? Here, we investigated the consistency of using one single Period-Luminosity Relation (PLR) for Cepheids in both the anchor and calibrator boxes.

As well described above with the procedure and the analysis, in a standard scenario, the Cepheids in the anchor and calibrator boxes, and SNe Ia are all fitted together with one PLR throughout the anchor and calibrator boxes. We can also perform individual fitting where we build the PLR for Cepheids in the anchor box to which we accurately know the distance. We then use this PLR information of slope and intercept in the calibrator box by fixing them while fitting for the distance modulus for galaxies in the calibrator box. And we confirm that we obtain similar results in both procedures. In the SH0ES analysis, PLR breaks are well studied, but those PLR breaks are based on the pulsation period of Cepheids (P > or < 10 days). In this work, we focus on a PLR break with distance.

We allow the Cepheid PLR intercept to change after a certain distance, which would affect the inference of distance moduli to affected galaxies, now requiring two different absolute magnitudes, $M_{B1}$ and $M_{B2}$ as fit parameters to fit for unaffected and affected SN distances in the calibrator box. We found that $M_{B2}$ is preferred by the data and is more than 3.5 $\sigma$ away from $M_{B1}$. We also found that in the case of the PLR break/$G$-transition hypothesis, we obtain a best-fit value of $H_0 = 65.90 \pm 1.70$~\text{km/s/Mpc}, which is in excellent agreement with the best-fit value of the Hubble constant as inferred from CMB data~\cite{Planck:2018vyg}, thereby mitigating the Hubble Tension.

\subsection{Novel Features of This Analysis}

Previous research has explored transitions in local physics as a potential solution to the Hubble tension. Perivolaropoulos and Skara \cite{Perivolaropoulos:2022khd} conducted a reanalysis of the SH0ES data, maintaining a constant $M_W^H$ and permitting variations in the absolute magnitudes of Type Ia supernovae ($M_{B1}$ and $M_{B2}$) before and after the transition. Their findings indicated that this method could somewhat ease the tension but did not completely resolve it. Meanwhile, Wojtak and Hjorth \cite{Wojtak:2022bct} investigated the color parameter $\beta$ in Type Ia supernovae and discovered notable differences between calibration and cosmological samples. Their results suggested that the local $H_0$ measurement is influenced by the chosen SN reference color, and by modifying this reference color, they were able to align the local $H_0$ value with the Planck measurement..

Building on these insights, our analysis introduces three novel features compared to previous studies:
\begin{enumerate}
    \item It incorporates the physics of a gravitational transition, enforcing a physically motivated transition of the Cepheid Period-Luminosity Relation (PLR) intercept at the same distance (time) as the transition of the SNe Ia absolute magnitude.
    \item It enforces this transition on $M_W^H$ rather than simply allowing it, providing a more natural assumption in the context of a gravitational transition. Both $M_W^H$ and $M_B$ are expected to change. Future analyses could consider connecting both with $\Delta G / G$, making $\Delta G / G$ the only new parameter to fit.
    \item  It allows for the SNe Ia standardized peak luminosity to vary with Chandrasekhar mass as $L \propto M_c^n$, where $n$ is the scaling index and is free to take both positive and negative values as a derived parameter.
    
\end{enumerate}

Using this approach and fixing $\Delta G / G \simeq 0.04$, which is well consistent with nucleosynthesis constraints, we find that the Hubble tension is naturally and fully resolved. The results obtained lead to a best-fit value of $H_0$ that is more constrained and significantly more consistent with the Planck value compared to previous studies, even without including a Planck prior or Planck data, as done in the case of Early Dark Energy (EDE) models.

\subsection{Summary of Findings}

First, we started by reconstructing the distance ladder as done in the SH0ES analysis. Then we allowed the Cepheid PLR intercept to change by implementing an increase in $G$ at a distance within the calibrator box. We observed that a Cepheid PLR intercept shift after a certain distance is equally preferred as assuming a single PLR. To incorporate the Cepheid PLR break with distance, we allowed a $G$-transition between 7 - 40~Mpc in the distance ladder. We found that a $G$-transition at a look-back distance of $\mu_t=32$ (approx. 25 Mpc), with an effective gravitational constant that was stronger in the past by an amount $\Delta G/G_{\textrm{N}} = 4\%$ (PLR intercept shift of 0.09 units by using Eq.~\ref{eq:PLR_intercept_G_1}), is equally preferred by the observational data from Cepheids and type Ia SNe as the hypothesis of no $G$-transition.

In performing our fit, we allowed for the SNe Ia standardized peak luminosity to vary with Chandrasekhar mass as $L \propto M_c^n$, where $n$ is the scaling index. We inferred a best-fit value of $n = -4.24 \pm 1.06$, whose inverse scaling is in agreement with the theoretical prediction $n=-0.97$ of Wright and Li~\cite{Wright2018}, which used a semi-analytic model for SNe light curves.

Taken together, our results provide circumstantial evidence for a cosmologically recent $G$-transition as a resolution to the Hubble tension. Unlike the proposal of a $G$-transition at 40~Mpc as suggested in \cite{Marra:2021fvf} as a resolution to the Hubble tension, where the authors assumed that SN standardized peak luminosity $L$ scales in proportion to the Chandrasekhar mass $M_c$, our scenario suggests an \textit{inverse} relationship between $L$ and $M_c$ in line with the expectations of~\cite{Wright2018}.

\subsection{Future Directions}

The analysis can be improved by incorporating the latest SH0ES data~\cite{Riess:2022mme}, which includes the luminosity-metallicity parameter as discussed by Perivolaropoulos and Skara in \cite{Perivolaropoulos:2022khd}, while retaining the novel features of the present analysis.

Furthermore, the dependence of all parameters involved in the analysis, including the period-luminosity parameter $b_W$, its intercept $M^H_W$, and the luminosity-metallicity parameter $Z_W$, on $\Delta G/G$ could be included. A fit could be made for the additional parameter $\Delta G/G$ and the transition distance or redshift. This would introduce a model dependence but provide a more physically motivated approach.

In our analysis, we assumed a precise value $B=4$ for the parameter $B$, which determines the correction to the intercept of the Cepheid light curves (Eq.~\ref{eq:lum_change}) in the presence of a $G$-transition. The true value for a given Cepheid was found to vary between $B = 3.46$ and $B = 4.52$, depending on the number of crossings of the Cepheid across the instability strip in the HR diagram~\cite{Sakstein2019}. Our value should thus be interpreted as an assumed average value for all Cepheids in our study. However, in order to estimate the error of this choice on our inferred value of the Hubble constant, we have also performed our analysis for $\mu_t = 32$ and $\Delta G/G_{\textrm{N}} = 4\%$ with the extreme values of $B = 3.46$ and $B = 4.52$, obtaining values of ($H_0 = 66.58 \pm 1.64$ \text{km/s/Mpc} and $n =-3.91 \pm 1.06$) and ($H_0 = 65.01 \pm 1.44$ \text{km/s/Mpc} and $n = -4.78 \pm 0.93$), respectively.

Recent studies have shown that a 5\% change at the 2-$\sigma$ level between the early universe and present-day is allowed, such as Planck 2018 CMB data combined with BAO data~\cite{Wang:2020bjk, Ballardini:2021evv}, and results from BBN~\cite{Alvey:2019ctk}. Studies have also suggested and constrained time variations of $G$ in the late universe (see, e.g.,~\cite{Mould:2014iga}). A more recent study \cite{Lamine:2024xno} confirms that a positive increase in Gravitational Constant $G$ is consistent by CMB + BAO DESI observations. Considering these hints from the extremely low redshift of the observational datasets and the 5-$\sigma$ tension between Planck and SH0ES results, we believe one should question and stress-test assumptions for the standard distance ladder. Our work found that the data do not disfavor the possibility of PLR breaks with distance. This work also suggests that one should further examine the SNe Ia luminosity-Mass ($M_c$) relations with upcoming data. Changing these two relations can completely change the result of the inference of the standard distance ladder and help explain the Hubble Tension.

\section*{Acknowledgements}
We thank Himansh Rathore, Shouvik Roy Choudhary, and Vikram Rentala for their insightful discussions. We acknowledge IUCAA, Pune, India, for the use of their computational facilities. The authors want to thank IFPU, Trieste, for supporting the visit where a part of the work was done. Authors R and AM are supported by "Theoretical Astroparticle Physics" (TAsP), iniziativa specifica INFN.
The work of R and A.M.  was partially supported by the research grant number 2022E2J4RK ``PANTHEON: 
Perspectives in Astroparticle and Neutrino THEory with Old and New messengers'' under the program PRIN 2022 funded by the Italian Ministero dell’Universit\`a 
e della Ricerca (MUR).
 This article is based upon work from COST Action CA21136 - Addressing observational tensions in cosmology with systematics and fundamental physics (CosmoVerse), supported by COST (European Cooperation in Science and Technology). This project was also supported by the Hellenic Foundation for Research and Innovation (H.F.R.I.), under the "First call for H.F.R.I. Research Projects to support Faculty members and Researchers and the procurement of high-cost research
equipment Grant" (Project Number: 789).

\appendix

\bibliographystyle{JHEP.bst}
\bibliography{references}

\providecommand{\href}[2]{#2}\begingroup\raggedright\begin{thebibliography}{10}

\bibitem{Wright2018}
B.S.~Wright and B.~Li, \emph{{Type Ia supernovae, standardizable candles, and
  gravity}}, \href{https://doi.org/10.1103/PhysRevD.97.083505}{\emph{Phys. Rev.
  D} {\bfseries 97} (2018) 083505}.

\bibitem{Planck:2018vyg}
{\scshape Planck} collaboration, \emph{{Planck 2018 results. VI. Cosmological
  parameters}},
  \href{https://doi.org/10.1051/0004-6361/201833910}{\emph{Astron. Astrophys.}
  {\bfseries 641} (2020) A6}
  [\href{https://arxiv.org/abs/1807.06209}{{\ttfamily 1807.06209}}].

\bibitem{Riess:2021jrx}
A.G.~Riess et~al., \emph{{A Comprehensive Measurement of the Local Value of the
  Hubble Constant with 1 km/s/Mpc Uncertainty from the Hubble Space Telescope
  and the SH0ES Team}},
  \href{https://doi.org/10.3847/1538-4357/ac5c5b}{\emph{Astrophys. J.}
  {\bfseries 934} (2022) L7}
  [\href{https://arxiv.org/abs/2112.04510}{{\ttfamily 2112.04510}}].

\bibitem{Freedman:2019jwv}
W.L.~{Freedman}, B.F.~{Madore}, D.~{Hatt}, T.J.~{Hoyt}, I.S.~{Jang},
  R.L.~{Beaton} et~al., \emph{{The Carnegie-Chicago Hubble Program. VIII. An
  Independent Determination of the Hubble Constant Based on the Tip of the Red
  Giant Branch}}, \href{https://doi.org/10.3847/1538-4357/ab2f73}{\emph{\apj}
  {\bfseries 882} (2019) 34}
  [\href{https://arxiv.org/abs/1907.05922}{{\ttfamily 1907.05922}}].

\bibitem{Yuan:2019npk}
W.~Yuan, A.G.~Riess, L.M.~Macri, S.~Casertano and D.~Scolnic, \emph{{Consistent
  Calibration of the Tip of the Red Giant Branch in the Large Magellanic Cloud
  on the Hubble Space Telescope Photometric System and a Re-determination of
  the Hubble Constant}},
  \href{https://doi.org/10.3847/1538-4357/ab4bc9}{\emph{Astrophys. J.}
  {\bfseries 886} (2019) 61}
  [\href{https://arxiv.org/abs/1908.00993}{{\ttfamily 1908.00993}}].

\bibitem{Soltis:2020gpl}
J.~Soltis, S.~Casertano and A.G.~Riess, \emph{{The Parallax of $\omega$
  Centauri Measured from Gaia EDR3 and a Direct, Geometric Calibration of the
  Tip of the Red Giant Branch and the Hubble Constant}},
  \href{https://doi.org/10.3847/2041-8213/abdbad}{\emph{Astrophys. J. Lett.}
  {\bfseries 908} (2021) L5}
  [\href{https://arxiv.org/abs/2012.09196}{{\ttfamily 2012.09196}}].

\bibitem{Freedman:2020dne}
W.L.~Freedman et~al., \emph{{Calibration of the Tip of the Red Giant Branch
  (TRGB)}}, \href{https://doi.org/10.3847/1538-4357/ab7339}{\emph{Astrophys.
  J.} {\bfseries 891} (2020) 57}
  [\href{https://arxiv.org/abs/2002.01550}{{\ttfamily 2002.01550}}].

\bibitem{Efstathiou:2013via}
G.~Efstathiou, \emph{{H0 Revisited}},
  \href{https://doi.org/10.1093/mnras/stu278}{\emph{Mon. Not. Roy. Astron.
  Soc.} {\bfseries 440} (2014) 1138}
  [\href{https://arxiv.org/abs/1311.3461}{{\ttfamily 1311.3461}}].

\bibitem{Cardona:2016ems}
W.~Cardona, M.~Kunz and V.~Pettorino, \emph{{Determining $H_0$ with Bayesian
  hyper-parameters}},
  \href{https://doi.org/10.1088/1475-7516/2017/03/056}{\emph{JCAP} {\bfseries
  03} (2017) 056} [\href{https://arxiv.org/abs/1605.06008}{{\ttfamily
  1605.06008}}].

\bibitem{Zhang:2017aqn}
B.~Zhang et~al., \emph{{A blinded determination of $H_0$ from low-redshift Type
  Ia supernovae, calibrated by Cepheid variables}},
  \href{https://doi.org/10.1093/mnras/stx1504}{\emph{Mon. Not. Roy. Astron.
  Soc.} {\bfseries 471} (2017) 2254}
  [\href{https://arxiv.org/abs/1706.07573}{{\ttfamily 1706.07573}}].

\bibitem{Feeney:2017sgx}
S.M.~Feeney et~al., \emph{{Prospects for resolving the Hubble constant tension
  with standard sirens}},
  \href{https://doi.org/10.1103/PhysRevLett.122.061105}{\emph{Phys. Rev. Lett.}
  {\bfseries 122} (2019) 061105}
  [\href{https://arxiv.org/abs/1707.00007}{{\ttfamily 1707.00007}}].

\bibitem{Dhawan:2017ywl}
S.~Dhawan, S.W.~Jha and B.~Leibundgut, \emph{{Measuring the Hubble constant
  with Type Ia supernovae as near-infrared standard candles}},
  \href{https://doi.org/10.1051/0004-6361/201731501}{\emph{Astron. Astrophys.}
  {\bfseries 609} (2018) A72}
  [\href{https://arxiv.org/abs/1707.00715}{{\ttfamily 1707.00715}}].

\bibitem{Follin:2017ljs}
B.~Follin and L.~Knox, \emph{{Insensitivity of The Distance Ladder Hubble
  Constant Determination to Cepheid Calibration Modeling Choices}},
  \href{https://doi.org/10.1093/mnras/stx1784}{\emph{Mon. Not. Roy. Astron.
  Soc.} {\bfseries 477} (2018) 4534}
  [\href{https://arxiv.org/abs/1707.01175}{{\ttfamily 1707.01175}}].

\bibitem{Riess:2018kzi}
A.G.~Riess et~al., \emph{{Milky Way Cepheid Standards for Measuring Cosmic
  Distances and Application to Gaia DR2: Implications for the Hubble
  Constant}}, \href{https://doi.org/10.3847/1538-4357/aac82e}{\emph{Astrophys.
  J.} {\bfseries 861} (2018) 126}
  [\href{https://arxiv.org/abs/1804.10655}{{\ttfamily 1804.10655}}].

\bibitem{Shanks:2018rka}
T.~Shanks, L.M.~Hogarth and N.~Metcalfe, \emph{{Gaia Cepheid parallaxes and
  'Local Hole' relieve $H_0$ tension}},
  \href{https://doi.org/10.1093/mnras/sty3240}{\emph{Mon. Not. Roy. Astron.
  Soc.} {\bfseries 484} (2019) L64}
  [\href{https://arxiv.org/abs/1810.02595}{{\ttfamily 1810.02595}}].

\bibitem{Huang:2019yhh}
C.~Huang et~al., \emph{{Hubble constant tension between CMB lensing and BAO
  measurements}}, \href{https://doi.org/10.1093/mnras/stz2875}{\emph{Mon. Not.
  Roy. Astron. Soc.} {\bfseries 490} (2019) 3793}
  [\href{https://arxiv.org/abs/1908.04281}{{\ttfamily 1908.04281}}].

\bibitem{Riess:2020fzl}
A.G.~Riess et~al., \emph{{The Accuracy of the Hubble Constant Measurement
  Verified Through Cepheid Amplitudes}},
  \href{https://doi.org/10.3847/1538-4357/abe1c5}{\emph{Astrophys. J.}
  {\bfseries 908} (2021) L6}
  [\href{https://arxiv.org/abs/2012.08534}{{\ttfamily 2012.08534}}].

\bibitem{Poulin:2018cxd}
V.~Poulin, T.L.~Smith, T.~Karwal and M.~Kamionkowski, \emph{{Early Dark Energy
  Can Resolve The Hubble Tension}},
  \href{https://doi.org/10.1103/PhysRevLett.122.221301}{\emph{Phys. Rev. Lett.}
  {\bfseries 122} (2019) 221301}
  [\href{https://arxiv.org/abs/1811.04083}{{\ttfamily 1811.04083}}].

\bibitem{Evslin:2017qdn}
J.~Evslin, A.A.~Sen and Ruchika, \emph{{Price of shifting the Hubble
  constant}}, \href{https://doi.org/10.1103/PhysRevD.97.103511}{\emph{Phys.
  Rev. D} {\bfseries 97} (2018) 103511}
  [\href{https://arxiv.org/abs/1711.01051}{{\ttfamily 1711.01051}}].

\bibitem{DiValentino:2021izs}
E.~Di~Valentino, O.~Mena, S.~Pan, L.~Visinelli, W.~Yang, A.~Melchiorri et~al.,
  \emph{{In the realm of the Hubble tension\textemdash{}a review of
  solutions}}, \href{https://doi.org/10.1088/1361-6382/ac086d}{\emph{Class.
  Quant. Grav.} {\bfseries 38} (2021) 153001}
  [\href{https://arxiv.org/abs/2103.01183}{{\ttfamily 2103.01183}}].

\bibitem{Bernal:2016gxb}
J.L.~Bernal, L.~Verde and A.G.~Riess, \emph{{The trouble with $H_0$}},
  \href{https://doi.org/10.1088/1475-7516/2016/10/019}{\emph{JCAP} {\bfseries
  10} (2016) 019} [\href{https://arxiv.org/abs/1607.05617}{{\ttfamily
  1607.05617}}].

\bibitem{Verde:2019ivm}
L.~Verde, T.~Treu and A.G.~Riess, \emph{{Tensions between the Early and the
  Late Universe}},
  \href{https://doi.org/10.1038/s41550-019-0902-0}{\emph{Nature Astron.}
  {\bfseries 3} (2019) 891} [\href{https://arxiv.org/abs/1907.10625}{{\ttfamily
  1907.10625}}].

\bibitem{Knox:2019rjx}
L.~Knox and M.~Millea, \emph{{Hubble constant hunter\textquoteright{}s guide}},
  \href{https://doi.org/10.1103/PhysRevD.101.043533}{\emph{Phys. Rev. D}
  {\bfseries 101} (2020) 043533}
  [\href{https://arxiv.org/abs/1908.03663}{{\ttfamily 1908.03663}}].

\bibitem{Dutta:2018vmq}
K.~Dutta, Ruchika, A.~Roy, A.A.~Sen and M.M.~Sheikh-Jabbari, \emph{{Beyond
  $\Lambda $CDM with low and high redshift data: implications for dark
  energy}}, \href{https://doi.org/10.1007/s10714-020-2665-4}{\emph{Gen. Rel.
  Grav.} {\bfseries 52} (2020) 15}
  [\href{https://arxiv.org/abs/1808.06623}{{\ttfamily 1808.06623}}].

\bibitem{Akarsu:2021fol}
O.~Akarsu, S.~Kumar, E.~\"Oz\"ulker and J.A.~Vazquez, \emph{{Relaxing
  cosmological tensions with a sign switching cosmological constant}},
  \href{https://doi.org/10.1103/PhysRevD.104.123512}{\emph{Phys. Rev. D}
  {\bfseries 104} (2021) 123512}
  [\href{https://arxiv.org/abs/2108.09239}{{\ttfamily 2108.09239}}].

\bibitem{Marra:2021fvf}
V.~Marra and L.~Perivolaropoulos, \emph{{Rapid transition of Geff at zt
  $\ensuremath{\simeq}0.01$ as a possible solution of the Hubble and growth
  tensions}}, \href{https://doi.org/10.1103/PhysRevD.104.L021303}{\emph{Phys.
  Rev. D} {\bfseries 104} (2021) L021303}
  [\href{https://arxiv.org/abs/2102.06012}{{\ttfamily 2102.06012}}].

\bibitem{Perivolaropoulos:2022khd}
L.~Perivolaropoulos and F.~Skara, \emph{{Challenges for \(\Lambda\)CDM: An
  update}},
  \href{https://doi.org/10.1016/j.astropartphys.2022.102677}{\emph{Astropart.
  Phys.} {\bfseries 136} (2022) 102677}
  [\href{https://arxiv.org/abs/2105.05208}{{\ttfamily 2105.05208}}].

\bibitem{Wojtak:2022bct}
R.~Wojtak and J.~Hjorth, \emph{{Intrinsic tension in the supernova sector of
  the local Hubble constant measurement and its implications}},
  \href{https://doi.org/10.1093/mnras/stac1878}{\emph{Mon. Not. Roy. Astron.
  Soc.} {\bfseries 515} (2022) 2790}
  [\href{https://arxiv.org/abs/2206.08160}{{\ttfamily 2206.08160}}].

\bibitem{Ruchika:2023ugh}
Ruchika, H.~Rathore, S.~Roy~Choudhury and V.~Rentala, \emph{{A gravitational
  constant transition within cepheids as supernovae calibrators can solve the
  Hubble tension}},
  \href{https://doi.org/10.1088/1475-7516/2024/06/056}{\emph{JCAP} {\bfseries
  06} (2024) 056} [\href{https://arxiv.org/abs/2306.05450}{{\ttfamily
  2306.05450}}].

\bibitem{Riess:2016jrr}
A.G.~Riess et~al., \emph{{A 2.4\% Determination of the Local Value of the
  Hubble Constant}},
  \href{https://doi.org/10.3847/0004-637X/826/1/56}{\emph{Astrophys. J.}
  {\bfseries 826} (2016) 56}
  [\href{https://arxiv.org/abs/1604.01424}{{\ttfamily 1604.01424}}].

\bibitem{GaiaEDR3}
A.G.A.~Brown, A.~Vallenari, T.~Prusti, J.H.J.~de~Bruijne, C.~Babusiaux,
  M.~Biermann et~al., \emph{{Gaia Early Data Release 3}},
  \href{https://doi.org/10.1051/0004-6361/202039657e}{\emph{{Astronomy and
  Astrophysics}} {\bfseries 650} (2021) C3}.

\bibitem{Pietrzynski:2019jed}
G.~Pietrzyński et~al., \emph{{A distance to the Large Magellanic Cloud that is
  precise to one per cent}},
  \href{https://doi.org/10.1038/s41586-019-0999-4}{\emph{Nature} {\bfseries
  567} (2019) 200} [\href{https://arxiv.org/abs/1903.08096}{{\ttfamily
  1903.08096}}].

\bibitem{Humphreys:2013eja}
E.M.L.~Humphreys et~al., \emph{{Toward a New Distance to the Active Galaxy NGC
  4258. II. H$_2$O Maser Absolute Proper Motions and Accelerations}},
  \href{https://doi.org/10.1088/0004-637X/775/1/13}{\emph{Astrophys. J.}
  {\bfseries 775} (2013) 13} [\href{https://arxiv.org/abs/1307.6031}{{\ttfamily
  1307.6031}}].

\bibitem{Riess:2019cxk}
A.G.~Riess et~al., \emph{{Large Magellanic Cloud Cepheid Standards Provide a
  1\% Foundation for the Determination of the Hubble Constant and Stronger
  Evidence for Physics beyond $\Lambda$CDM}},
  \href{https://doi.org/10.3847/1538-4357/ab1422}{\emph{Astrophys. J.}
  {\bfseries 876} (2019) 85}
  [\href{https://arxiv.org/abs/1903.07603}{{\ttfamily 1903.07603}}].

\bibitem{Fitzpatrick:1999by}
E.L.~Fitzpatrick, \emph{{Correcting for the effects of interstellar
  extinction}}, \href{https://doi.org/10.1086/316293}{\emph{Publ. Astron. Soc.
  Pac.} {\bfseries 111} (1999) 63}.

\bibitem{ForemanMackey:2012ig}
D.~Foreman-Mackey, D.W.~Hogg, D.~Lang and J.~Goodman, \emph{{emcee: The MCMC
  Hammer}}, \href{https://doi.org/10.1086/670067}{\emph{Publ. Astron. Soc.
  Pac.} {\bfseries 125} (2013) 306}
  [\href{https://arxiv.org/abs/1202.3665}{{\ttfamily 1202.3665}}].

\bibitem{Riess:2022mme}
A.G.~Riess, L.~Breuval, W.~Yuan, S.~Casertano, L.M.~Macri, J.B.~Bowers et~al.,
  \emph{{Cluster Cepheids with High Precision Gaia Parallaxes, Low Zero-point
  Uncertainties, and Hubble Space Telescope Photometry}},
  \href{https://doi.org/10.3847/1538-4357/ac8f24}{\emph{Astrophys. J.}
  {\bfseries 938} (2022) 36}
  [\href{https://arxiv.org/abs/2208.01045}{{\ttfamily 2208.01045}}].

\bibitem{book_edi}
A.S.~Eddington, \emph{The Internal Constitution of the Stars}, Cambridge
  University Press (1926).

\bibitem{ritter}
A.~{Ritter}, \emph{{Untersuchungen {\"u}ber die H{\"o}he der Atmosph{\"a}re und
  die Constitution gasf{\"o}rmiger Weltk{\"o}rper}},
  \href{https://doi.org/10.1002/andp.18792440910}{\emph{Annalen der Physik}
  {\bfseries 244} (1879) 157}.

\bibitem{martin}
C.~Martin and H.C.~Plummer, \emph{{On the Short-Period Variable RR Lyræ.
  (Plate 20, 21.)}},
  \href{https://doi.org/10.1093/mnras/75.7.566}{\emph{Monthly Notices of the
  Royal Astronomical Society} {\bfseries 75} (1915) 566}
  [\href{https://arxiv.org/abs/https://academic.oup.com/mnras/article-pdf/75/7/566/18494284/mnras75-0566.pdf}{{\ttfamily
  https://academic.oup.com/mnras/article-pdf/75/7/566/18494284/mnras75-0566.pdf}}].

\bibitem{edi18}
A.S.~Eddington, \emph{{On the Pulsations of a Gaseous Star and the Problem of
  the Cepheid Variables. Part I}},
  \href{https://doi.org/10.1093/mnras/79.1.2}{\emph{Monthly Notices of the
  Royal Astronomical Society} {\bfseries 79} (1918) 2}
  [\href{https://arxiv.org/abs/https://academic.oup.com/mnras/article-pdf/79/1/2/2833686/mnras79-0002.pdf}{{\ttfamily
  https://academic.oup.com/mnras/article-pdf/79/1/2/2833686/mnras79-0002.pdf}}].

\bibitem{edi19}
A.S.~{Eddington}, \emph{{The Pulsations of a Gaseous Star and the Problem of
  the Cepheid Variables.: Part II.}},
  \href{https://doi.org/10.1093/mnras/79.3.177}{\emph{\mnras} {\bfseries 79}
  (1919) 171}.

\bibitem{Sakstein2019}
J.~{Sakstein}, H.~{Desmond} and B.~{Jain}, \emph{{Screened fifth forces
  mediated by dark matter-baryon interactions: Theory and astrophysical
  probes}}, \href{https://doi.org/10.1103/PhysRevD.100.104035}{\emph{\prd}
  {\bfseries 100} (2019) 104035}
  [\href{https://arxiv.org/abs/1907.03775}{{\ttfamily 1907.03775}}].

\bibitem{Paxton_2019}
B.~Paxton, R.~Smolec, J.~Schwab, A.~Gautschy, L.~Bildsten, M.~Cantiello et~al.,
  \emph{Modules for experiments in stellar astrophysics ({MESA}): Pulsating
  variable stars, rotation, convective boundaries, and energy conservation},
  \href{https://doi.org/10.3847/1538-4365/ab2241}{\emph{The Astrophysical
  Journal Supplement Series} {\bfseries 243} (2019) 10}.

\bibitem{Pinto}
P.A.~Pinto and R.G.~Eastman, \emph{{The Type Ia supernova width luminosity
  relation}},  \href{https://arxiv.org/abs/astro-ph/0006171}{{\ttfamily
  astro-ph/0006171}}.

\bibitem{woosley}
S.E.~Woosley and T.A.~Weaver, \emph{The physics of supernova explosions},
  \href{https://doi.org/10.1146/annurev.aa.24.090186.001225}{\emph{Annual
  Review of Astronomy and Astrophysics} {\bfseries 24} (1986) 205}
  [\href{https://arxiv.org/abs/https://doi.org/10.1146/annurev.aa.24.090186.001225}{{\ttfamily
  https://doi.org/10.1146/annurev.aa.24.090186.001225}}].

\bibitem{amendola}
L.~{Amendola}, S.~{Corasaniti} and F.~{Occhionero}, \emph{{Time variability of
  the gravitational constant and Type Ia supernovae}}, {\emph{arXiv e-prints}
  (1999) astro} [\href{https://arxiv.org/abs/astro-ph/9907222}{{\ttfamily
  astro-ph/9907222}}].

\bibitem{Gazt}
E.~Gazta\~naga, E.~Garc\'{\i}a-Berro, J.~Isern, E.~Bravo and I.~Dom\'{\i}nguez,
  \emph{Bounds on the possible evolution of the gravitational constant from
  cosmological type-ia supernovae},
  \href{https://doi.org/10.1103/PhysRevD.65.023506}{\emph{Phys. Rev. D}
  {\bfseries 65} (2001) 023506}.

\bibitem{CO1996}
B.W.~{Carroll} and D.A.~{Ostlie}, \emph{{An Introduction to Modern
  Astrophysics; Benjamin Cummings, 1996. ISBN 0-201-54730-9}} (1996).

\bibitem{Akaike}
H.~Akaike, \emph{A new look at the statistical model identification},
  \href{https://doi.org/10.1109/TAC.1974.1100705}{\emph{IEEE Transactions on
  Automatic Control} {\bfseries 19} (1974) 716}.

\bibitem{Schwarz:1978tp}
G.~Schwarz, \emph{{Estimating the Dimension of a Model}},
  \href{https://doi.org/10.1214/aos/1176344136}{\emph{Annals Statist.}
  {\bfseries 6} (1978) 461}.

\bibitem{Wang:2020bjk}
K.~Wang and L.~Chen, \emph{{Constraints on Newton\textquoteright{}s constant
  from cosmological observations}},
  \href{https://doi.org/10.1140/epjc/s10052-020-8137-x}{\emph{Eur. Phys. J. C}
  {\bfseries 80} (2020) 570}
  [\href{https://arxiv.org/abs/2004.13976}{{\ttfamily 2004.13976}}].

\bibitem{Ballardini:2021evv}
M.~Ballardini, F.~Finelli and D.~Sapone, \emph{{Cosmological constraints on the
  gravitational constant}},
  \href{https://doi.org/10.1088/1475-7516/2022/06/004}{\emph{JCAP} {\bfseries
  06} (2022) 004} [\href{https://arxiv.org/abs/2111.09168}{{\ttfamily
  2111.09168}}].

\bibitem{Alvey:2019ctk}
J.~Alvey, N.~Sabti, M.~Escudero and M.~Fairbairn, \emph{{Improved BBN
  Constraints on the Variation of the Gravitational Constant}},
  \href{https://doi.org/10.1140/epjc/s10052-020-7727-y}{\emph{Eur. Phys. J. C}
  {\bfseries 80} (2020) 148}
  [\href{https://arxiv.org/abs/1910.10730}{{\ttfamily 1910.10730}}].

\bibitem{Mould:2014iga}
J.~Mould and S.~Uddin, \emph{{Constraining a possible variation of G with Type
  Ia supernovae}}, \href{https://doi.org/10.1017/pasa.2014.9}{\emph{Publ.
  Astron. Soc. Austral.} {\bfseries 31} (2014) 15}
  [\href{https://arxiv.org/abs/1402.1534}{{\ttfamily 1402.1534}}].

\bibitem{Lamine:2024xno}
B.~Lamine, Y.~Ozdalkiran, L.~Mirouze, F.~Erdogan, S.~Ilic, I.~Tutusaus et~al.,
  \emph{{Cosmological measurement of the gravitational constant $G$ using the
  CMB}},  \href{https://arxiv.org/abs/2407.15553}{{\ttfamily 2407.15553}}.

\end{thebibliography}\endgroup

\end{document}